\title{\textbf{A study of convolution models for background correction of BeadArrays}}
\author{
        \large
        \tt {Rohmatul Fajriyah}
        \thanks{Universitas Islam Indonesia, Yogyakarta, Indonesia (\href{mailto:rfajriyah@fmipa.uii.ac.id}{rfajriyah@fmipa.uii.ac.id}). In 2012/2013: Graz University of Technology, Institute of Statistics (\href{mailto:fajriyah@student.tugraz.at}{fajriyah@student.tugraz.at}) Research supported by Austrian Science Fund (FWF), Project P24302-N18. }\\ 
        \vspace{-5mm}
}
\date{}

\documentclass[12pt]{article}
\usepackage[left=1.5in, right=1in, top=1in, bottom=0.75in, includefoot, headheight=13.6pt]{geometry}
\usepackage[colorlinks]{hyperref}
\usepackage{subfig,graphicx,amssymb,amsthm,amstext,amsmath,mathtools,mdwlist,color,float,rotating,rotfloat,caption,xtab,setspace,relsize,exscale}
\usepackage[titletoc]{appendix}
\usepackage[square, comma, numbers, sort&compress]{natbib}
\usepackage[T1]{fontenc}
\numberwithin{equation}{section}

\numberwithin{figure}{section}
\usepackage{sectsty}

\allowdisplaybreaks

\begin{document}
\maketitle

\begin{abstract}
The RMA, since its introduction in \cite{Iri03a, Iri03b, Iri06}, has gained popularity among bioinformaticians. It has evolved from the exponential-normal convolution to the gamma-normal convolution, from single to two channels and from the Affymetrix to the Illumina platform.

The Illumina design has provided two probe types: the regular and the control probes. This design is very suitable for studying the probability distribution of both and one can apply the convolution model to compute the true intensity estimator. The availability of benchmarking data set at Illumina platform, the {\it Illumina spike-in}, helps researchers to evaluate their proposed method for Illumina BeadArrays.

In this paper, we study the existing convolution models for background correction of Illumina BeadArrays in the literature and give a new estimator for the true intensity, where the intensity value is exponentially or gamma distributed and the noise has lognormal distribution. We compare the performance of the models on the Illumina spike-in  data set, based on various criteria, for example, root and mean square error, $L_{1}$ error, Kullback-Leibler coefficient, and some adapted criteria from Affycomp \cite{Cop04}. We then provide a simulation study to measure the consistency of the error of background correction and the parametrization. We also study the performance of all models on the FFPE data set.

Our study shows that our GLN model (with the method of moments for parameter estimation) is the optimal one for the benchmarking data set with benchmarking criteria, while the gamma-normal model has the best performance for the benchmarking data set with simulation criteria.  At the public data set of FFPE, the gamma-normal and the exponential-gamma models with MLE cannot be used and our proposed models ELN and GLN have the best performance, showing a moderate error in background correction and in the parametrization.

\end{abstract}

\newpage{}
\pagenumbering{arabic}
\section{Introduction} \label{sec1}
There are various processes in producing data from microarray experiments and each process contributes a noise to the data. The noise can be of two types, biological and non-biological. Non-biological noise should be avoided or at least minimized.

Sources for the non-biological noises are, for example,  the chip itself, the scanner, or fluctuations in the electric network. Therefore, the data needs to be adjusted. The pre-processing will adjust the intensity value (\cite{Hub05a,Hub05b}) and provides the estimation of the true intensity.

To estimate the intensity value, researchers proposed additive and multiplicative models and also {\it additive-muliplicative error models}, see e.g.\ Huber et al.\cite{Hub04}. In case of additive models, the underlying distribution is generally chosen as normal (log-normal), exponential, or gamma-$t$ mixture in the parametric approach ( \cite{All09},  \cite{Bol03}, \cite{Che11}, \cite{Hoc06}, \cite{Iri03a, Iri03b, Iri06}, and \cite{Pla11,Pla12}).

Irizarry et al. \cite{Iri03a, Iri03b, Iri06} and Bolstad et al. \cite{Bol03}, on the Affymetrix platform, have estimated the true intensity values based on the convolution model in the background correction step of their robust multi-array average (RMA) pre-processing method. They assumed that the true intensity $S$ is exponentially distributed and the background noise $B$ is normally distributed.

Plancade et al. \cite{Pla11, Pla12} showed that the RMA model (in \cite{Bol03} and \cite{Iri03a, Iri03b, Iri06}) does not fit Illumina BeadArrays: using the exponential-normal convolution leads to a large distance between the observed and the modeled intensities. They proposed, instead, the implementation of gamma distribution for the intensity value and normal distribution for the noise.

The simulation study of Plancade et al. \cite{Pla11, Pla12} showed that the gamma-normal model performs better than the existing  exponential-normal convolution model, giving a more accurate and correct fit for the observed intensities in Illumina BeadArray.

Using gamma distribution for the intensity values in Illumina BeadArray has been first suggested by Xie et al. \cite{Xie09}.

The studies of Baek et al. \cite{Bae07} (in the background correction of the image processing) and Chen et al. \cite{Che11} show that the noise distribution is usually skewed in different degrees. In their studies, based on  simulated and real data sets, Baek et al. \cite{Bae07} conclude that the gamma distribution is well suited for the noise. It accounts for the intensities with a positive lower bound and is very flexible in its shape, including asymmetric exponential type and symmetric normal type.

The proposed convolution of exponential-gamma distribution by Chen et al. \cite{Che11} improves the intensity estimation and the detection of differentially expressed genes in the case when the intensity to noise ratio is large and the noise has a skewed distribution.

In view of the remarks above, it is natural to model both the true intensity and the background noise in Illumina BeadArray as gamma distributed.
In an earlier version of this paper we have developed an estimator for the true intensity based on the gamma-gamma convolution model of RMA. However, this model does not fit very well the Illumina benchmarking data set. Independently, Triche et al. \cite{Tri13} proposed and applied the gamma-gamma model to pre-process Illumina methylation arrays.

In this paper we introduce a new model for background correction in Illumina BeadArrays where the true intensity value is exponentially or gamma distributed and the noise has lognormal distribution. As we will see, this model avoids the difficulties with the gamma-gamma model and has an overall satisfactory performance.

We note that a new method reducing the bias of the maximum likelihood  estimator of the shape parameter of the gamma distribution was  proposed by Zhang \cite{Zha13}. But since our samples are very large, bias is not a problem in our studies.

 Our paper is organized as follows. In Section~\ref{sec2}  we describe the Illumina BeadArray technology and in Section~\ref{sec3} we review previous work related to the background correction for Illumina BeadArrays. Our model is described in Section~\ref{sec4}. Sections ~\ref{sec5} and ~\ref{sec6} are explaining the benchmarking and the simulation studies in Section~\ref{sec7} provide a performance comparison. Finally, Section~\ref{sec8} gives the conclusions and indications of future work.

\section{Illumina BeadArrays technology}\label{sec2}
Illumina technology is one of the most advanced technologies in analyzing gene ex- pression by microarrays. It can be used to profile partially degraded ribonucleic acid (RNA) which is usually found in FFPE samples, by the cDNA-mediated Annealing, Selection, Exten- sion, and Ligation (DASL) assays method.

The huge amount of available FFPE data makes the the Illumina platform very important because of the nature of the DASL assay method, which can deal with the partially degraded RNA to profile the gene expression on the samples.

\begin{figure}[!h]
\begin{center}
\includegraphics[width=4in,height=3.25in]{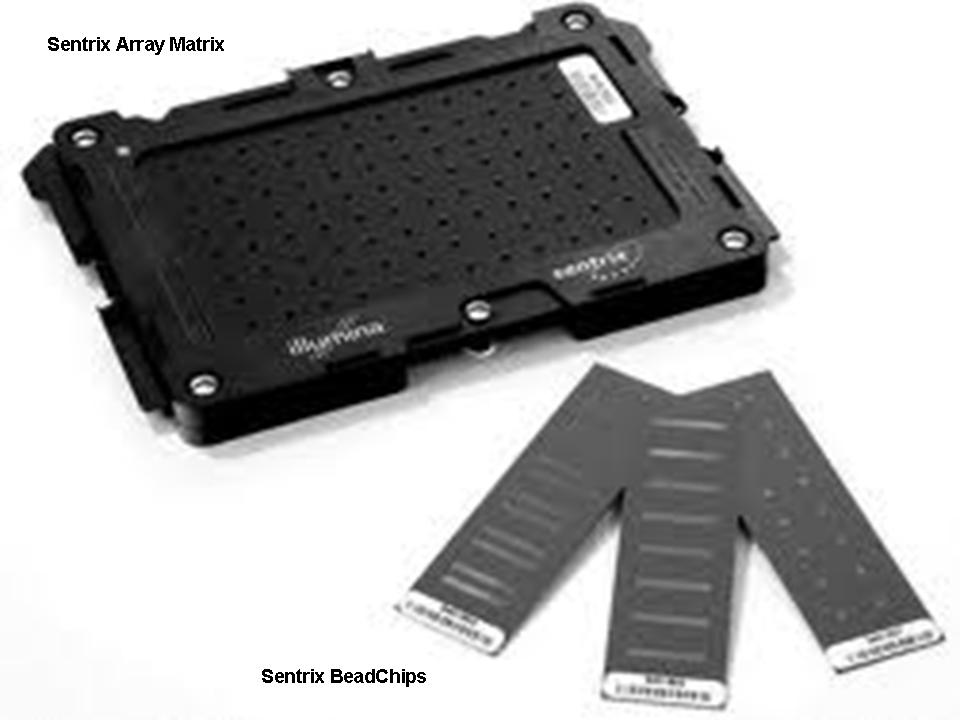}
\end{center}
\captionsetup{font={footnotesize}}
\caption{Illumina platforms, \cite{Fan06}} \label{fig123}
\end{figure}

\begin{figure}[!h]
\begin{center}
\includegraphics[width=5in,height=3.25in]{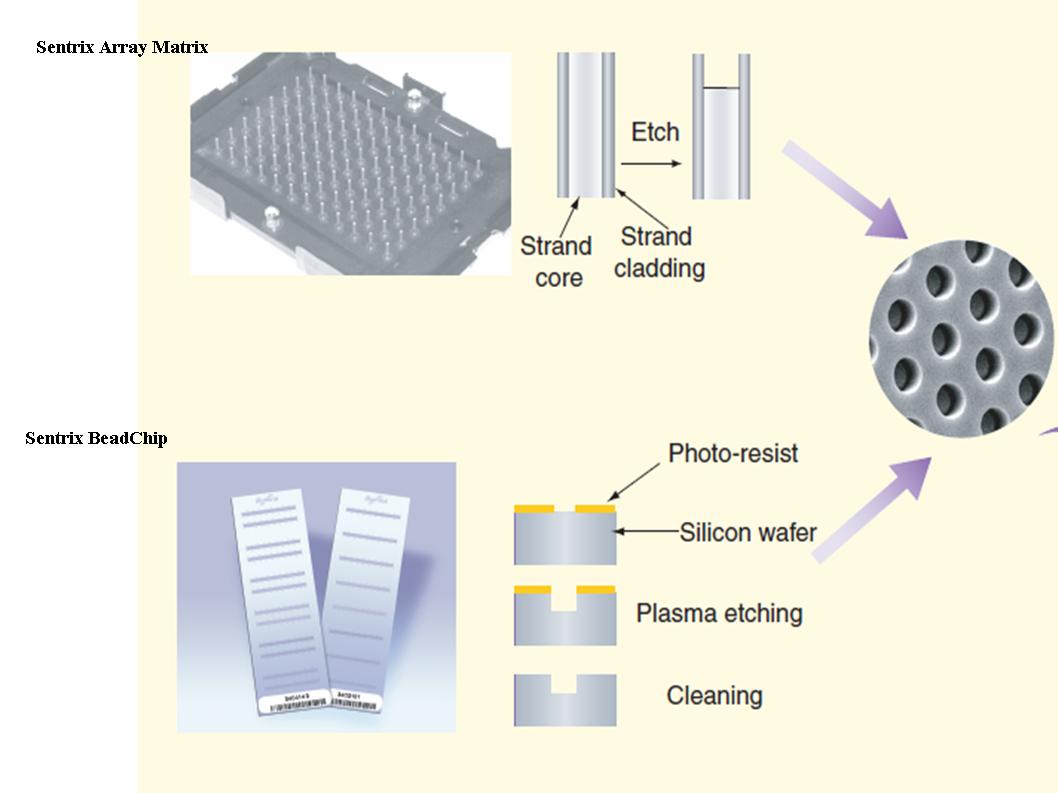}
\end{center}
\captionsetup{font={footnotesize}}
\caption{Pattern substrate of Illumina platform, \cite{Ste05}.} \label{fig124}
\end{figure}

The Illumina platform has small feature size, dense features and the ability to analyze multiple samples in parallel. Illumina provides two formats of microarrays (\cite{Fan05}, \cite{Fan06}, \cite{Mil09} and \cite{Ste05}), the $\text{Sentrix}^{\textregistered}$ Array Matrix (SAM) and the Sentrix BeadChip (SBC). See Figure \ref{fig123}. The pattern substrate can be seen at Figure \ref{fig124}.

The Array Matrix arranges fiber optic bundles, each containing 50,000 fibers within distance 5-$\mu$m, into an Array of Arrays$\texttrademark$ format of a 96-well microtiter plate. On one end of the fiber optic bundles, the core of each fiber is etched to form a nanowell for the 3-$\mu$m silica beads.

In the BeadChip format, one to several microarrays is arranged on silicon slides that have been processed by micro-electromechanical systems (MEMS) technology to also have nanowells that support the self assembly of beads.

$\\ \\$

\begin{figure}[!h]
\begin{center}
\includegraphics[width=3.5in,height=2.5in]{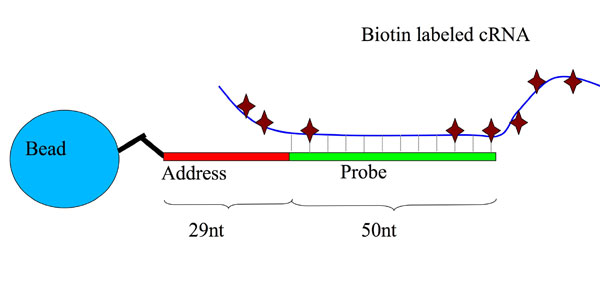}
\end{center}
\captionsetup{font={footnotesize}}
\caption{The design of Illumina bead. In this figure, the bead is shown to be coated by one oligonucleotide only. In the real bead, it is coated by hundreds of thousands of copies of a specific oligonucleotide.  \url{http://bitesizebio.com/articles/how-dna-microarrays-are-built/}, retrieved June 29, 2012.} \label{fig125}
\end{figure}

\begin{figure}[!h]
\begin{center}
\includegraphics[width=5in,height=3.5in]{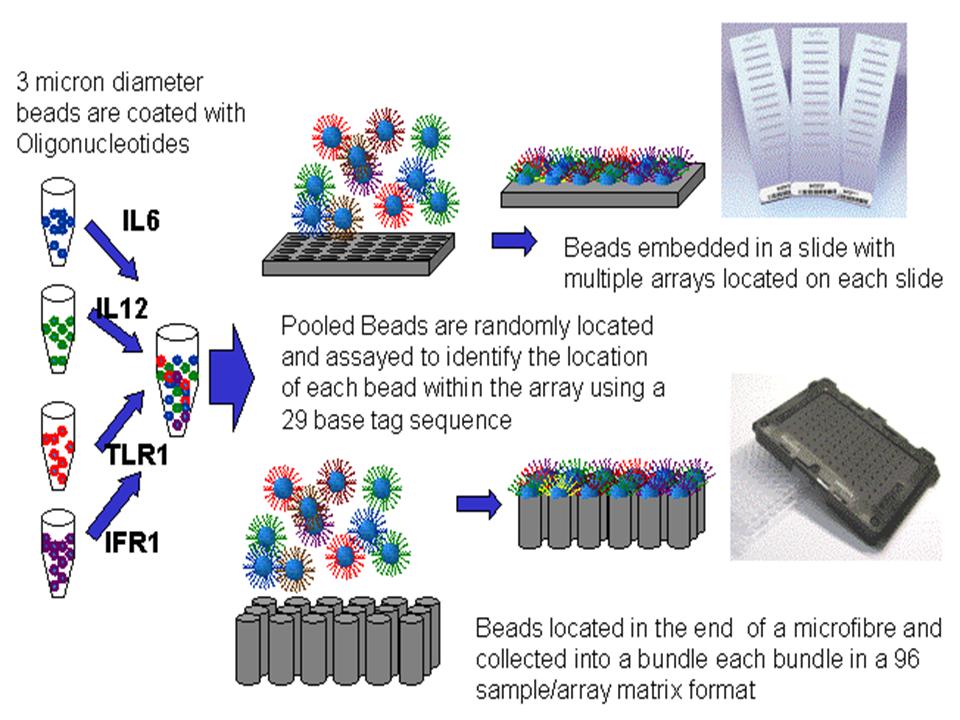}
\end{center}
\captionsetup{font={footnotesize}}
\caption{Production process of Illumina, \url{http://www.ipc.nxgenomics.org/newsletter/no8.htm}, retrieved June 29, 2012.} \label{fig126}
\end{figure}

Steemers and Gunderson \cite{Ste05} explain the three parts of Illumina arrays manufacturing (see Figure \ref{fig126}). The three parts are:
\begin{enumerate}
\item{The first part is the creation of a master bead pool consisting of 1,536-250,000 different bead types. For the quality control, it includes the negative control beads. Oligonucleotide capture probes are immobilized individually by bead type in a bulk process. Each bead type in an array comes from a single immobilization event, reducing array-to-array feature variability. The design of Illumina bead can be seen at Figure \ref{fig125}.}
\item{The second step is the random self assembly of the master pool of bead types into etched wells on the array substrate, where each bead type has an average 30 times representation - a strategy that provides the statistical accuracy of multiple measurements.}
\item{The third step is the identification of each bead on the array, through a decoding process. This process provides information of each bead and performs a quality control of the feature in every array.}
\end{enumerate}

$\\$

\section{Previous work}\label{sec3}
Affymetrix is the pioneer and most widely used platform for microarray gene expression experiments. The tools and algorithms to handle the data are numerous, both free and commercial. Some methods for pre-processing are available. Examples for the background correction step are: MAS5.0 by Affymetrix, multiplicative model based expression index (MMBE) by Li and Wong \cite{LiW01}, RMA in Irizarry et al. \cite{Iri03a, Iri03b, Iri06} and Bolstad et al. \cite{Bol03}, GC-RMA by Wu et al. \cite{Wu04} and maximum likelihood estimation based on the normal-exponential convolution model by Silver et al. \cite{Sil09}.

Illumina is one of the alternative platforms and is increasingly popular. A few statistical methods have been developed for BeadArray data and there is no consensus yet for the pre-processing steps \cite{Shi10}. Xie et al. \cite{Xie09} mention that for the background correction step, Illumina bead studio gives two options (no background correction and background substraction) and the packages for BeadArrays in R  provide three options (no background correction, background substraction and RMA background correction).

Ding et al. \cite{Din08} extended the RMA model by proposing the model-based background correction method (MBCB) and showed that their model leads to a more precise determination of the gene expression and a better biological interpretation of Illumina BeadArray data.

The studies of Chen et al. \cite{Che11} and Plancade et al. \cite{Pla11,Pla12} show that their background correction models are made by adapting the RMA Affymetrix model. As Forcheh et al. \cite{For12}, pointed out, most preprocessing methods for Illumina bead arrays are ported from the Affymetrix microarray platform.

\subsection{Background correction by RMA} \label{sec31}
In modeling the intensity values, the RMA (\cite{Bol03}, and \cite{Iri03a,Iri03b,Iri06}) assumes that the intensity values are affected by the noises of the chip. The RMA model is as follows:
 \begin{equation} \label{eq31}
P=S+B
\end{equation}
where
$P=PM$ is the observed probe level intensity of perfect match probes,
$S$ is the true signal, with $S \sim f_{1} (s ;  \theta) = \text{Exp}(\theta), \theta >0$, and
$B$ is the background noise of the chip with $B \sim f_{2} (b ; \mu,\sigma ^{2}) = \mathcal{N}(\mu,\sigma ^{2}), \mu \in \mathbb{R}, \sigma^{2} >0$. To avoid negative intensity values, we truncate $B$ at 0 from below; this will not change its density function  $f_{2} (b ; \mu,\sigma ^{2})$ for $b>0$.

Assuming independence, the joint density of the two-dimensional random variable $(S,B)$ is
\begin{eqnarray} \label{eq32}
f_{S,B} (s, b;\mu,\sigma^{2},\theta)=\frac{1}{\theta}\exp \Bigg \{ -\frac{s}{\theta}\Bigg \} f_{2}(b;\mu,\sigma^{2}),s>0\, .
\end{eqnarray}
Furthermore, the transformation formula for two-dimensional densities gives that joint density of $S$ and $P$ is
\begin{equation} \label{eq33}
f_{S,P} (s,p ; \mu,\sigma^{2},\theta)=\frac{\exp \Big\{\frac{\sigma^{2}}{2\theta^{2}}-
\frac{(p-\mu)}{\theta}\Big\}}{\theta}f_{2}\left(s;p-\mu-\frac{\sigma^{2}}{\theta},\sigma^{2}\right)\, , \quad 0< s< p.
\end{equation}

From equation (\ref{eq33}) we get the marginal density of $P$ and the conditional density of $S$ given $P$ in equations (\ref{eq34}) and (\ref{eq35}) below, respectively:
\begin{equation} \label{eq34}
\frac {\exp\Big\{\frac{\sigma^{2}}{2\theta^{2}}-\frac{(p-\mu)}{\theta}\Big\}}{\theta} \left(\Phi \left( \frac{\mu_{S.P}}{\sigma} \right)+\Phi \left( \frac{p-\mu_{S.P}}{\sigma} \right)  -1\right)
\end{equation}
\begin{eqnarray} \label{eq35}
\frac{f_2 (s; \mu_{S, P}, \sigma^2)}{\left(\Phi \left( \frac{\mu_{S.P}}{\sigma} \right)+\Phi \left( \frac{p-\mu_{S.P}}{\sigma} \right)  -1\right)}
\end{eqnarray}
where $\mu_{S.P}=p-\mu-\frac{\sigma^{2}}{\theta}$.

The background adjusted intensity is computed by the estimated signal given the observed intensity. It is the conditional expectation
$$
E(S\mid P=p)
=\frac{1}{\left(\Phi \left( \frac{\mu_{S.P}}{\sigma} \right)+\Phi \left( \frac{p-\mu_{S.P}}{\sigma} \right)  -1\right)}\int_0^p f_{2}\left(s;\mu_{S.P},\sigma^{2}\right)ds.
$$
The substitution $s=\mu_{S, P}+\sigma t$ yields
$$\int_0^p s f_2 (s; \mu_{S, P}, \sigma^2)=\mu_{S, P} \left(\Phi\left(\frac{p-\mu_{S,P}}{\sigma}\right)+\Phi\left(\frac{\mu_{S,P}}{\sigma}\right)-1\right) +\sigma \left(\phi\left(\frac{\mu_{S,P}}{\sigma}\right) -\phi\left(\frac{p-\mu_{S,P}}{\sigma}\right)\right)$$
and thus the estimator is written as \cite{Bol03}
\begin{eqnarray} \label{eq36}
 \mu_{S.P}+\sigma \frac{\phi(\frac{\mu_{S.P}}{\sigma}) - \phi(\frac{p-\mu_{S.P}}{\sigma})}{\Phi(\frac{\mu_{S.P}}{\sigma}) +\Phi(\frac{p-\mu_{S.P}}{\sigma})-1}.
\end{eqnarray}

Note that modelling the noise as a truncated normal variable has the consequence that the noise equals 0 with a positive
probability $p_0$, a rather unpleasant feature of the model. As pointed out in \cite{Xie09}, however, in practical cases $p_0$ is rather small, so this problem can be disregarded. To avoid this difficulty, one can model the noise as the absolute value of an $\mathcal{N}(\mu,\sigma ^{2})$ variable, which changes the calculations above. However, since in this paper we will provide a background correction model fitting reality considerably better, we do not give the details here.

\subsection{Exponential-normal MBCB} \label{sec32}

Xie et al. \cite{Xie09} use the same underlying distributions in Equation (\ref{eq31}) in estimating the true intensity value. The difference with the RMA (\cite{Bol03}, and \cite{Iri03a,Iri03b,Iri06}) are
\begin{enumerate}
\item Xie et al. \cite{Xie09} use $+\infty$ as the upper bound of the integral to compute the marginal density function and the conditional expectation of the true intensity value. On the other hand, RMA uses $p$ as the upper bound of the integration.

The estimator of the true intensity value of Xie et al. \cite{Xie09} is
\begin{eqnarray} \label{eq37}
 \mu_{S.P}+\sigma \frac{\phi(\frac{\mu_{S.P}}{\sigma})}{\Phi(\frac{\mu_{S.P}}{\sigma})}.
\end{eqnarray}

\item Under the convolution model (\ref{eq31}), where the true intensity value is assumed exponentially distributed and the noise is normally distributed, we need to estimate the parameters $\theta, \mu,$ and $\sigma^{2}$.  Xie et al. \cite{Xie09}  offer three parameter estimation methods: method of moments, maximum likelihood and bayesian. On the other hand, RMA applies the \emph{ad-hoc} method.
\end{enumerate}

\subsection{Gamma-normal convolution} \label{sec33}

Plancade et al. \cite{Pla11,Pla12} introduced gamma-normal convolution to model the background correction of Illumina BeadArray. The model is based on the RMA background correction of Affymetrix GeneChip. Plancade et al. \cite{Pla11, Pla12} assume that the intensity value is gamma distributed and the noise is normally distributed.

Under the model background correction in (\ref{eq31}), $f_{P}$ is the convolution of $f_{S}$ and $f_{B}$. The true intensity $S$ is estimated by the conditional expectation of $S$ given $P=p$:
\begin{eqnarray} \label{eq38}
\tilde{S}(p)
&=& \frac{\int s f^{\text{gam}}_{\alpha,\beta}(s) f^{\text{norm}}_{\mu,\sigma}(p-s) \mathrm{d}s }{\int f^{\text{gam}}_{\alpha,\beta}(s) f^{\text{norm}}_{\mu,\sigma}(p-s) \mathrm{d}s}
\end{eqnarray}
where
$$
f_{\alpha, \beta}^{\text{gam}} (x;\alpha, \beta)=\frac{\beta^{\alpha} x^{\alpha-1} e^{ -\beta x}} {\Gamma (\alpha)}, \quad \alpha, \beta, x>0
$$
is the gamma density.
When $S$ is gamma distributed and $B$ is normally distributed, then equation (\ref{eq38}) has no analytic expression like (\ref{eq36}). Plancade et al. \cite{Pla11,Pla12} implemented the Fast Fourier Transform (\emph{fft}) to estimate the parameter. Moreover, equation (\ref{eq38}) can be written as
\begin{eqnarray} \label{eq39}
\tilde{S}(p\mid \Theta)
&=& \frac{\alpha \beta \int f^{\text{gam}}_{\alpha+1,\beta}(s) f^{\text{norm}}_{\mu,\sigma}(p-s) \mathrm{d}s }{\int f^{\text{gam}}_{\alpha,\beta}(s) f^{\text{norm}}_{\mu,\sigma}(p-s) \mathrm{d}s}\, ,
\end{eqnarray}

where $\Theta=(\mu,\sigma,\alpha,\beta)$, $S$ and $B$ are independent and $sf_{\alpha,\beta}^{gam}(s)=\alpha \beta f_{\alpha+1,\beta}^{gam}(s)$ is valid for every $s > 0$.

\subsection{Exponential-Gamma convolution} \label{sec34}
Chen et al. \cite{Che11} proposed for the distribution of the true intensity and its noise, under the convolution model of Equation (\ref{eq31}), the exponential and gamma distribution, respectively. Therefore, $\\$
$S \sim f_{1} (s ;  \theta) = \text{Exp}(\theta)$  $\\$
and $\\$
$B \sim f (b ; \alpha, \beta) = \text{GAM}(\alpha, \beta)$, \\
where $s, b, \theta, \alpha, \beta >0$.

The corrected background intensity for the proposed model (\cite{Che11}) is
\begin{eqnarray} \label{eq310}
\hat{S}
&=& p -\frac{\int_0^p  b^\alpha e^{-(\frac{1}{\beta} -\frac{1}{\theta})b} db}{ \int_0^p  b^{\alpha-1} e^{-(\frac{1}{\beta} -\frac{1}{\theta}) b} db}.
\end{eqnarray}

\section{Results} \label{sec4}
\subsection{Exponential-lognormal convolution}\label{sec41}

\subsubsection{Estimation of the true intensity value}\label{sec411}

Consider the model (\ref{eq31}), when the true intensity $S$ is  exponentially distributed, i.e.\  $S \sim f_{1} (s ;  \theta) = \theta e^{-\theta s}, \theta, s  >0$, and the background noise $B$ is  lognormally distributed, $B \sim f_{2} (b ; \mu,\sigma^{2}) = \frac{e^{-\frac{(\ln b -\mu)^{2}}{2\sigma^{2}}}}{b \sigma\sqrt{2\pi}}, \mu \in \mathbb{R}, \sigma^{2}, b >0$.
The joint density function of $S$ and $B$ equals
$$f_{S,B}(s,b)=\theta e^{-\theta s} \frac{e^{-\frac{(ln b -\mu)^{2}}{2\sigma^{2}}}}{b \sigma\sqrt{2\pi}},$$
and thus the joint density function of $S$ and $P$ is
$$f_{S,P}(s,p)=\theta e^{-\theta s}\frac{e^{-\frac{(\ln (p-s) -\mu)^{2}}{2\sigma^{2}}}}{(p-s) \sigma\sqrt{2\pi}}.$$
Consequently, the marginal density function of $P$ equals
\begin{eqnarray}
f_{P}(p)
&=& \mathlarger{\int \limits_{0}^{p} f_{S,P}(s,p) ds} \nonumber \\
&=& \mathlarger{\int \limits_{0}^{p} \theta e^{-\theta s} \frac{e^{-\frac{(\ln (p-s) -\mu)^{2}}{2\sigma^{2}}}}{(p-s) \sigma\sqrt{2\pi}} ds}. \nonumber
\end{eqnarray}
Using the substitution $\ln (p-s)=z$ we can evaluate the last integral as follows:
\begin{eqnarray} \label{eq12}
f_{P}(p)
&=& \mathlarger{\int \limits_{-\infty}^{\text{ln}p} \frac{\theta e^{-\theta (p-e^{z})}  e^{-\frac{(z -\mu)^{2}} {2\sigma^{2}}} }{\sigma\sqrt{2\pi}} dz} \nonumber \\
&=& \mathlarger{   \frac{\theta e^{-\theta p}}{ \sigma\sqrt{2\pi}}\int \limits_{-\infty}^{\ln p} e^{-\frac{(z -\mu)^{2}}{2\sigma^{2}}} \displaystyle \sum_{k=0}^{\infty} \frac{\theta^{k} e^{kz}}{k!} dz} \nonumber \\
&=& \mathlarger{   \frac{\theta e^{-\theta p}}{\sigma\sqrt{2\pi}} \displaystyle \sum_{k=0}^{\infty} \frac{\theta ^{k} }{k!} \int \limits_{-\infty}^{\ln p} e^{-\frac{(z -\mu)^{2}}{2\sigma^{2}}+kz}  dz} \nonumber \\
&=& \mathlarger{ \frac{\theta e^{-\theta p}}{\sigma\sqrt{2\pi}} \displaystyle \sum_{k=0}^{\infty} \frac{\theta ^{k}e^{k(\mu+\frac{k\sigma ^{2}}{2})} }{k!} \int \limits_{-\infty}^{\ln p} e^{-\frac{(z -(\mu+k\sigma^{2}))^{2}}{2\sigma^{2}}} dz} \nonumber \\
&=&\theta e^{-\theta p} \displaystyle \sum_{k=0}^{\infty} \frac{\theta^{k}}{k!}e^{k(\mu+\frac{k}{2}\sigma^{2})}\Phi \Big(\frac{\ln p - (\mu+k\sigma^{2})}{\sigma}\Big) \nonumber \\
&=&\theta e^{-\theta p} C_{a},
\end{eqnarray}
where
$$C_{a}=\displaystyle \sum_{k=0}^{\infty} \frac{\theta^{k}}{k!}e^{k(\mu+\frac{k}{2}\sigma^{2})}\Phi \Big(\frac{\ln p - (\mu+k\sigma^{2})}{\sigma}\Big).$$
The conditional density function of $S$ under $P=p$ is now obtained as
\begin{eqnarray} \label{eq13}
f_{S\mid P}(s \mid p)
&=&\frac{f_{S,P}(s,p)}{f_{P}(p)} \nonumber \\
&=&\frac{\theta e^{-\theta s}\frac{e^{-\frac{(\ln (p-s) -\mu)^{2}}{2\sigma^{2}}}}{(p-s) \sigma\sqrt{2\pi}}} {\theta e^{-\theta p} C_{a}}\nonumber \\
&=&\frac{e^{\theta (p-s)} e^{-\frac{(\ln (p-s) -\mu)^{2}}{2\sigma^{2}}}}{C_{a}(p-s) \sigma\sqrt{2\pi}}.
\end{eqnarray}
The true intensity value is estimated by taking the expectation of the conditional density function in (\ref{eq13}). It is computed as follows:
\begin{eqnarray}
E(S \mid P=p)
&=&\mathlarger{\int \limits_{0}^{p} s f(s \mid p) ds} \nonumber \\
&=&\frac{e^{\theta p}}{C_{a}} \mathlarger{\int \limits_{0}^{p} \frac{se^{-\theta s} e^{-\frac{(\ln (p-s) -\mu)^{2}}{2\sigma^{2}}}} {(p-s) \sigma\sqrt{2\pi}} ds}. \nonumber
\end{eqnarray}
Using the substitution $\ln (p-s)=z$ we see that the last integral equals
\begin{eqnarray} \label{eq15}
&=&\frac{p}{C_{a}} \mathlarger{\int \limits_{-\infty}^{\ln p} \frac{(1-\frac{e^{z}}{p})e^{\theta e^{z}} e^{-\frac{(z -\mu)^{2}}{2\sigma^{2}}}} {\sigma\sqrt{2\pi}} dz}\nonumber \\
&=&\frac{p}{C_{a}} \Bigg[ \mathlarger{\int \limits_{-\infty}^{\ln p} \frac{e^{-\frac{(z -\mu)^{2}}{2\sigma^{2}}}}{\sigma\sqrt{2\pi}} e^{\theta e^{z}}dz - \int \limits_{-\infty}^{\ln p} \frac{e^{-\frac{(z -\mu)^{2}}{2\sigma^{2}}}}{\sigma\sqrt{2\pi}}\frac{e^{z}}{p} e^{\theta e^{z}}dz}\Bigg] \nonumber \\
&=&\frac{p}{C_{a}} \Bigg[ C_{a} - \mathlarger{\int \limits_{-\infty}^{\ln p} \frac{e^{-\frac{(z -\mu)^{2}}{2\sigma^{2}}}}{\sigma\sqrt{2\pi}}\frac{e^{z}}{p} e^{\theta e^{z}}dz}\Bigg] \nonumber \\
&=&p-\frac{e^{\mu+\frac{\sigma^{2}}{2}}}{C_{a}}\mathlarger{\int \limits_{-\infty}^{\ln p} \frac{e^{-\frac{(z -(\mu+\sigma^{2}))^{2}}{2\sigma^{2}}}e^{\theta e^{z}}}{\sigma\sqrt{2\pi}} dz} \nonumber \\
&=&p-\frac{e^{\mu+\frac{\sigma^{2}}{2}}}{C_{a}} \displaystyle \sum_{k=0}^{\infty} \frac{\theta^{k}}{k!}e^{k(\mu+\frac{k+2}{2}\sigma^{2})} \mathlarger {\int \limits_{-\infty}^{\ln p} \frac{e^{-\frac{(z -(\mu+(k+1)\sigma^{2}))^{2}}{2\sigma^{2}}}} {\sigma\sqrt{2\pi}} dz} \nonumber \\
&=&p-\frac{e^{\mu+\frac{\sigma^{2}}{2}}}{C_{a}} \displaystyle \sum_{k=0}^{\infty} \frac{\theta^{k}}{k!}e^{k(\mu+\frac{k+2}{2}\sigma^{2})}\Phi \Big (\frac{\ln p - (\mu+(k+1)\sigma^{2})}{\sigma} \Big)\nonumber\\
&=&p-\frac{e^{\mu+\frac{\sigma^{2}}{2}}C_{b}}{C_{a}}
\end{eqnarray}
where
$$
C_{b}=\displaystyle \sum_{k=0}^{\infty} \frac{\theta^{k}}{k!}e^{k(\mu+\frac{k+2}{2}\sigma^{2})}\Phi \Big (\frac{\ln p - (\mu+(k+1)\sigma^{2})}{\sigma} \Big).
$$

\subsubsection{Parameter estimation} \label{ss16}

To estimate the parameters $\theta, \mu,$ and $\sigma$ in (\ref{eq15}), we can use various methods.

\begin{enumerate}
\item{Maximum likelihood (MLE)} \\
This is implemented by applying the \emph{optim} function to the likelihood function
\begin{eqnarray} \label{eq16}
L=\prod \theta e^{-\theta p} C_{a} \prod \frac{e^{-\frac{(\ln b -\mu)^{2}}{2\sigma^{2}}}}{b \sigma\sqrt{2\pi}}.
\end{eqnarray}
The log-likelihood function is
\begin{eqnarray*} \label{eq16}
l&=&\sum \Big \{ \ln (\theta) -\theta p +\ln (C_{a}) \Big\} +\sum \Big \{ -\frac{(\ln b -\mu)^{2}}{2\sigma^{2}} -\ln (b) -\ln (\sigma)-\frac{ \ln (2 \pi)}{2} \Big\}. \nonumber \\
\end{eqnarray*}

\item{Method of moments} \\
The method of moments is implemented as follows:
\begin{enumerate}
\item{Compute $E(S)=E(P)-E(B)$ and $\theta$ is estimated by $\frac{1} {E(S)}$}
\item{Compute $\mu$ and $\sigma$ from $E(\text{ln}(B))$ and $\sqrt{\text{var}(\text{ln}(B))}$ respectively}
\end{enumerate}
 and
\item{Plug-in} \\
The plug in estimation is implemented by estimating
\begin{enumerate}
\item{$\theta$ from regular probe intensities $P$ through MLE}
\item{$\mu$ and $\sigma$} from control probe intensities $B$ through MLE
\end{enumerate}
\end{enumerate}

\subsection{Gamma-lognormal convolution}\label{sec42}
\subsubsection{Estimation of the true intensity value}\label{sec421}
Consider now model (\ref{eq31}), when the true intensity $S$ is assumed to be gamma distributed, $S \sim f_{1} (s ;  \alpha, \beta) = \frac{ s^{\alpha-1}e^{-\frac{s}{\beta}}}{\beta^{\alpha} \Gamma(\alpha)}, \alpha, \beta, s  >0$, and the background noise $B$ is lognormally distributed,
i.e.\ $B \sim f_{2} (b ; \mu,\sigma ^{2}) = \frac{e^{-\frac{(\ln b -\mu)^{2}}{2\sigma^{2}}}}{b \sigma\sqrt{2\pi}}, \mu \in \mathbb{R}, \sigma^{2} >0$.
The joint density function of $S$ and $B$ is
\begin{equation} \label{eq48}
f_{S,B}(s,b)=\frac{ s^{\alpha-1}e^{-\frac{s}{\beta}}}{\beta^{\alpha} \Gamma(\alpha)}\frac{e^{-\frac{(\ln b -\mu)^{2}}{2\sigma^{2}}}}{b \sigma\sqrt{2\pi}}
\end{equation}
and thus joint density function of $S$ and $P$ is
\begin{equation} \label{eq49}
f_{S,P}(s,p)=\frac{ s^{\alpha-1}e^{-\frac{s}{\beta}}}{\beta^{\alpha} \Gamma(\alpha)} \frac{e^{-\frac{(\ln (p-s) -\mu)^{2}}{2\sigma^{2}}}}{(p-s) \sigma\sqrt{2\pi}}.
\end{equation}
Hence the marginal density function of $P$ is obtained as
\begin{eqnarray} \label{eq410}
f_{P}(p)
&=& \mathlarger {\int \limits_{0}^{p} f_{S,P}(s,p) ds} \nonumber \\
&=& \mathlarger {\int \limits_{0}^{p} \frac{ s^{\alpha-1}e^{-\frac{s}{\beta}}}{\beta^{\alpha} \Gamma(\alpha)} \frac{e^{-\frac{(\ln (p-s) -\mu)^{2}}{2\sigma^{2}}}}{(p-s) \sigma\sqrt{2\pi}} ds}.
\end{eqnarray}

Using the substitution $\ln (p-s)=z$, we get
\begin{eqnarray} \label{eq411}
f_{P}(p)
&=& \mathlarger {\int \limits_{\infty}^{\ln p} \frac{ p^{\alpha-1}(1-\frac{e^{z}}{p})^{\alpha-1} e^{-\frac{p}{\beta}} e^{\frac{ e^{z}}{\beta} } e^{-\frac{(z -\mu)^{2}}{2\sigma^{2}}} }{\beta^{\alpha} \Gamma(\alpha)\sigma\sqrt{2\pi}} dz}\nonumber \\
&=& \frac{p^{\alpha-1}e^{-\frac{p}{\beta}}}{\beta^{\alpha}\Gamma(\alpha) \sigma\sqrt{2\pi}} \mathlarger {\int \limits_{\infty}^{\ln p} e^{-\frac{(z -\mu)^{2}}{2\sigma^{2}}}
(1-\frac{e^{z}}{p})^{\alpha-1} e^{\frac{ e^{z}}{\beta} }dz}\nonumber \\
&=&\frac{p^{\alpha-1}e^{-\frac{p}{\beta}}}{\beta^{\alpha}\Gamma(\alpha)} \displaystyle \sum_{k=0}^{\infty}  \frac{(-1)^{k} \binom{\alpha-1}{k} }{p^{k}} \Bigg [ \sum_{n=0}^{\infty} \frac{e^{(k+n)(\mu +(k+n)\frac{\sigma^{2}}{2})}\Phi \Big( \frac{\ln p -(\mu+(k+n)\sigma^{2})}{\sigma}\Big)}{\beta^{n} n!} \Bigg ]\nonumber \\
&=&\frac{p^{\alpha-1}e^{-\frac{p}{\beta}}C_{c}}{\beta^{\alpha}\Gamma(\alpha)},
\end{eqnarray}

where
$$C_{c}=\displaystyle \sum_{k=0}^{\infty}   \sum_{n=0}^{\infty} \frac{(-1)^{k} \binom{\alpha-1}{k} }{p^{k}} \frac{e^{(k+n)(\mu +(k+n)\frac{\sigma^{2}}{2})}\Phi \Big( \frac{\ln p -(\mu+(k+n)\sigma^{2})}{\sigma}\Big)}{\beta^{n} n!}.$$
The conditional density function of $S$  under $P=p$ is now obtained as
\begin{eqnarray} \label{eq412}
f_{S\mid P}(s \mid p)
&=&\frac{f_{S,P}(s,p)}{f_{P}(p)}\nonumber \\
&=&\frac{\frac{\beta^{\alpha} s^{\alpha-1}e^{-\beta s}}{\Gamma(\alpha)} \frac{e^{-\frac{(\ln (p-s) -\mu)^{2}}{2\sigma^{2}}}}{(p-s) \sigma\sqrt{2\pi}}}{\frac{p^{\alpha-1}e^{-\frac{p}{\beta}}C_{c}}{\beta^{\alpha}\Gamma(\alpha)}}\nonumber \\
&=&\frac{e^{\frac{p}{\beta}}}{C_{c}p^{\alpha-1}}\frac{s^{\alpha-1}e^{-\frac{s}{\beta}} e^{-\frac{(\ln (p-s) -\mu)^{2}}{2\sigma^{2}}}} {(p-s) \sigma\sqrt{2\pi}}.
\end{eqnarray}

The true intensity value is estimated by taking the expectation of the conditional density function in (\ref{eq412}) It is computed as follows:
\begin{eqnarray}
E(S \mid P=p)
&=&\int_{0}^{p} s f(s \mid p) ds\nonumber \\
&=&\frac{e^{\frac{p}{\beta}}}{C_{c}p^{\alpha-1}} \int_{0}^{p} \frac{s^{\alpha}e^{-\frac{s}{\beta}} e^{-\frac{(\ln (p-s) -\mu)^{2}}{2\sigma^{2}}}} {(p-s) \sigma\sqrt{2\pi}} ds. \nonumber
\end{eqnarray}
Substituting  $\ln (p-s)=z$ the integral above becomes
\begin{eqnarray} \label{eq413}
&=&\frac{p}{C_{c}} \int_{-\infty}^{\ln p} \frac{e^{-\frac{(z -\mu)^{2}}{2\sigma^{2}}}}{\sigma\sqrt{2\pi}} (1-\frac{e^{z}}{p})^{\alpha} e^{\frac{e^{z}}{\beta}}dz\nonumber \\
&=&\frac{p}{C_{c}} \displaystyle \sum_{k=0}^{\infty}  \frac{(-1)^{k} \binom{\alpha}{k} }{p^{k}} \Bigg [ \sum_{n=0}^{\infty} \frac{e^{(k+n)(\mu +(k+n)\frac{\sigma^{2}}{2})}\Phi \Big( \frac{\ln p -(\mu+(k+n)\sigma^{2})}{\sigma}\Big)}{\beta^{n} n!} \Bigg ]\nonumber \\
&=&\frac{pC_{d}}{C_{c}},
\end{eqnarray}
where
$$C_{d}=\displaystyle \sum_{k=0}^{\infty}  \sum_{n=0}^{\infty} \frac{(-1)^{k} \binom{\alpha}{k} }{p^{k}}   \frac{e^{(k+n)(\mu +(k+n)\frac{\sigma^{2}}{2})}\Phi \Big( \frac{\ln p -(\mu+(k+n)\sigma^{2})}{\sigma}\Big)}{\beta^{n} n!}.$$

\subsubsection{Parameter estimation} \label{sec426}
To estimate the parameters $\alpha, \beta, \mu,$ and $\sigma$ in (\ref{eq413}), we can use either of the following methods.

\begin{enumerate}
\item{Maximum likelihood (MLE)} \\
This is implemented by applying the \emph{optim} function to the likelihood function
\begin{eqnarray} \label{eq414}
L=\prod \frac{p^{\alpha-1}e^{-\frac{p}{\beta}}C_{c}}{\beta^{\alpha}\Gamma(\alpha)}\prod \frac{e^{-\frac{(\ln b -\mu)^{2}}{2\sigma^{2}}}}{b \sigma\sqrt{2\pi}}.
\end{eqnarray}
The log-likelihood function is
\begin{eqnarray*} \label{eq415}
l&=&\sum \Big \{ \ln (C_{c}+(\alpha-1) \ln (p) -\frac{p}{\beta} - \alpha \ln (\beta)-\ln (\Gamma(\alpha))  \Big\} \nonumber \\
& & +\sum \Big \{ -\frac{(\ln b -\mu)^{2}}{2\sigma^{2}} -\ln (b) -\ln (\sigma)-\frac{\ln (2 \pi)}{2}  \Big\}.
\end{eqnarray*}

\item{Method of moments} \\
The method of moments is implemented as follows:
\begin{enumerate}
\item{Compute $\bar{S} = \bar{P} - \bar{B}$ and $S_{S}^{2}=S_{P}^{2}+S_{B}^{2}$ then $\alpha$ and $\beta$ are estimated by $\frac{\bar{S}}{\beta}$ and $\frac{S_{S}^{2}}{\bar{S}}$ respectively}
\item{Compute $\mu$ and $\sigma^{2}$ from $\bar{\ln (B)}$ and $S_{B}^{2}(\ln (B))$ respectively}
\end{enumerate}
 and
\item{Plug-in} \\
The plug-in estimation is implemented as follows:
\begin{enumerate}
\item{$\alpha$ and $\beta$ are estimated from regular probe intensities $P$ through MLE}
\item{$\mu$ and $\sigma^{2}$ are estimated from control probe intensities $B$ through MLE}
\end{enumerate}
\end{enumerate}

\section{Benchmarking} \label{sec5}
\subsection{Benchmarking data set}\label{sec51}

Illumina platform has provided a benchmarking data set, the Illumina spike-in \cite{Dun08a}.
These spike-in probes are targeting bacterial and viral genes absent from the mouse genome. These were added at specific concentrations on each sample. Therefore the change in expression level of a particular spike between samples is known a priori. The expression levels of the non-spikes should not change between samples.

There are twelve different concentrations of spike: 1000 picomolar (pM), 300 pM, 100 pM, 30 pM, 10 pM, 3 pM, 1 pM, 0.3 pM, 0.1 pM 0.03 pM, 0.01 pM and 0 pM.  It was replicated four times. Therefore, there are 48 samples and each sample has regular and control probes.

There are approximately about 48,000 probesets for each sample and in addition the 33 spike-in probes are added into it.  For the control probes, there are 1,616 probes. These control experiments are the benchmarking data sets of Illumina and are used to compare low-level analysis methods such as in Affymetrix platform.

\subsection{Benchmarking Criteria}\label{sec52}
We adopt the benchmarking criteria from Affymetrix platform, in the \emph{Affycomp} package, and some criteria which have been used by \cite{Xie09}, \cite{Che11}, and  \cite{Pla11,Pla12}.

Affycomp \cite{Cop04} has provided fourteen criteria and here we define different ranges for each classification. Cope et al. \cite{Cop04} defined the low,medium and high intensities as the nominal concentrations less than or equal to 2 pM, the nominal concentration between 4 and 32 pMs, and the nominal concentrations greater than or equal to 64 pM.

For the Illumina Spike-in data set, the high, medium and low concentration are defined respectively as, the nominal concentrations less than or equal to 1 pM, the nominal concentration between 3 and 30 pMs, and the nominal concentrations greater than or equal to 100 pM. Instead of using the slope, we used the $R^{2}$, because it is reflecting the best fit of the data: the quantity to measure how much percentage that the observed expressions is explained by the nominal concentrations.

 Criteria 1 to 9 below were computed by the author and  Criteria 10 to 14 were computed by implementing the \textit{assessSpikeIn2} and the \textit{assessSpikeIn} functions from the affycomp \textit{package}, with some adjustments. The benchmarking criteria are as follows:

\begin{enumerate}
\item \emph{Median SD}. It is believed that the variance of an expression measure across replicate arrays should be low, so the standard deviation (SD) will be low too, ideally zero. The median of the standard deviations across replicate arrays is chosen as the measurement due to its robustness.
\item \emph{Null log-fc IQR}. The non-spike in genes should not be differentially expressed across arrays. Therefore, the Inter-quartile range (IQR) of the log-fold-changes of the non- spike in genes is, ideally, zero.
\item \emph{Null log-fc 99.9\%}. As above but using the 99.9\% percentile.
\item \emph{Signal detects R2}. The R-squared (R2) is obtained from regressing the expression values on nominal concentrations in the spike-in data. The ideal value of R-squared is 1, because ideally an increment in the nominal concentration will be followed by an increment of the expression values, in the same scale.
\item \emph{Low.$R^2$}. This is obtained from regression of observed log concentration on nominal log concentration for genes with low intensities.
\item \emph{Med.$R^2$}. This is obtained from regression of observed log concentration on nominal log concentration for genes with medium intensities.
\item \emph{High.$R^2$}. As above but for genes with high intensities.
\item \emph{Obs-intended-fc $R^2$}. The $R^2$ that is obtained by regressing observed log-fold-changes against nominal log-fold-changes for the spike in genes.
\item \emph{Obs-(low)int-fc $R^2$}. The $R^2$ that is obtained by regressing observed log-fold-changes against nominal log-fold-changes for the spike in genes with low intensities.
\item \emph{Low AUC}. This is computed as the Area under the receiver operator characteristic (ROC) curve (up to 100 false positives) for genes with low intensities, and standardized. Therefore, the optimum value is 1.
\item \emph{Med AUC}. As above but for genes with medium intensities.
\item \emph{High AUC}. As above but for genes with high intensities.
\item \emph{Weighted avg AUC}. A weighted average of the previous 3 ROC curves with weights related to amount of data in each classification (low, medium and high).
\item \emph{All AUC}. An AUC for all intensities, 12 arrays.
\end{enumerate}

\subsection{Affycomp Plot}\label{sec53}
Affycomp contains some plots that are used as the supplemental support in the process of benchmarking against spike-in and dilution data set. Some of them are as in \cite{Cop04} and \url{http://affycomp.biostat.jhsph.edu}.

For the Illumina BeadArrays a slightly different usage is explained as follows:

\begin{enumerate}
\item{MA plot}\\
This plot uses 12 arrays representing a single experiment of Illumina spike-in and the fold changes are generated by comparing the first arrays in the set to each of the others. Spiked-in genes are symbolized by numbers representing the nominal $\log_{2}$ fold change for the gene. The non-spike-in genes with observed absolute fold changes larger than 2 are plotted in red. All other probe sets are represented with black.
\item{Variance across replicates plot}\\
Using the benchmarking data set, the variance of an expression measure across replicate array should be low. For each non-spiked-in gene in the arrays used in MA plot, the mean log expression and the observed standard deviation across the replicates are calculated. The resulting scatter plot is smoothed to generate a single curve representing mean standard deviation as a function of mean log expression. The standard deviation should be low and independent of expression level.
\item{Observed expression versus nominal expression plot}\\
In this plot the log observed intensity of spike-in gene is plotted against log  nominal concentration. The averaged values of observed intensities at each nominal concentration are used to produce a mean curve. Ideally, if the nominal concentration is doubled, so should be the observed intensity. Therefore, ideally the observed intensity should be linear in true concentrations with a slope of 1.
\item{ROC curve}\\
Identification of genes which are differentially expressed can be done by filtering the genes using a fold change exceeding a given threshold. An ROC curve offers a graphical representation of both specificity and sensitivity for such a rule. It is constructed by plotting the true positive (TP) rate (sensitivity) against false positive (FP) rate (1- specificity).
\item{Observed fold-change versus nominal fold-change}\\
The plotting of log fold-change observation and nominal is used for validation of differentially expressed genes.
\end{enumerate}

\subsection{Reproducibility}\label{sec54}
To assess which models reproduce best the benchmarking data, two measurements are applied (see e.g.\ Shamilov \cite{Sha06}):
\begin{enumerate}
\item{Root Mean Square Errors (RMSE),  RMSE$=\Big(\frac{\sum_{i=1}^{N} (y_{i}-x_{i})^{2}}{N}\Big)^{\frac{1}{2}}$}
\item{Kullback-Leibler (K-L), K-L$= \sum_{i=1}^{N} y_{i} \log \Big(\frac{y_{i}}{x_{i}}\Big)$, \\
where $x$ is the true intensity estimator and $y$ is the observed intensity}.
\end{enumerate}

\section{Simulation study} \label{sec6}

Let $N$ be the number of simulations, $n_{1}$ the sample size of regular probes, $n_{2}$ the sample size of negative probes, $\Theta$ is the original parameter vector of the underlying distribution in each model from the data set and $\hat{\Theta}$ is the parameter vector of the underlying distribution in each model from the simulation data.

The simulation is conducted by referring the convolution in Equation (\ref{eq31}), based on the underlying distribution. Once we have decided the model, then
\begin{enumerate}
\item{choose the parameters for the simulation. The parameters for the simulation are a combination of the minimum, median and maximum values of the original parameters. The original parameters are estimated from the data set based on the choosen model.}
\item{generate a sample for the true intensity ($S$), negative probes ($B$)  and regular probes ($S+B$)}
\item{estimate the parameters of the underlying distribution based on the generated sample ($\Theta$) and save}
\item{compute the estimation of the true intensity value ($\hat{S}$) and save}
\item{repeat the steps above $N$ times, then}
\item{compute the simulation criteria, to measure the bias of the background correction and the parameters:
\begin{enumerate}
\item{MSE$_{bc}$ is defined as \\MSE$_{bc} = \frac{1}{N} \displaystyle \sum_{l=1}^{N} \Big ( \frac{1}{n_{1}} \sum_{j=1}^{n_{1}} (\hat{S} (P_{j}^{l}| \hat{\Theta}_{l})-S_{j}^{l} )^{2} \Big )$}
\item{$L_{1}$ error is defined as \\ \\ $L_{1} = \frac{1}{N} \displaystyle \sum_{l=1}^{N} \frac{\mid \Theta-\hat{\Theta}_l\mid}  {\Theta} $}
\end{enumerate}}
\end{enumerate}

The lower the criteria the better the model would represent the right model for the data at hand.

\section{Performance studies} \label{sec7}
We compare all convolution models : Irizarry et al. \cite{Iri03a,Iri03b,Iri06} and Bolstad et al. \cite{Bol03}: RMA (Exponential-Normal), Plancade et al. \cite{Pla11,Pla12}:Gamma-Normal, Chen et al. \cite{Che11}: Exponential-Gamma, Xie et al. \cite{Xie09}: Exponential-Normal adjusted for Illumina BeadArrays with maximum likelihood estimation (MLE) for the parameters, Bayesian approach and the moment method, and the proposed models: exponential-lognormal and gamma-lognormal .

We will call the methods above, respectively, as follows: ENr, GN, EGm, ENm, ENm, ENn, ELNn, ELNm, ELNp, GLNn, GLNm, and GLNp. We use the MBCB \textit{package} (\cite{All09} and \cite{Xie09}) to adjust the intensity values of these existing models ENr, ENm, ENmc and ENn. Except that, the GN uses the NormalGamma \textit{package} (\cite{Pla11}).

The comparison is made for the benchmarking and non-benchmarking data set. For the non benchmarking data set, we use two of the formalin-fixed, paraffin-embedded (FFPE) data sets from Waldron et al. \cite{Wal12}: the FFPE of tumors from colorectal cancer patients (GSE32651, 1003 samples),  breast cancer metastases of the lymph node and autopsy tissues (GSE32490: GSE32489, 120 samples). Each sample has 24,526 probesets.

The link for the data set are \url{http://www.ncbi.nlm.nih.gov/geo/query/acc.cgi?acc=GSE32651} and \url{http://www.ncbi.nlm.nih.gov/geo/query/acc.cgi?acc=GSE32490}.

\subsection{Illumina Spike-in} \label{sec71}

\begin{table}[!h]
\begin{center}
\caption{Reproducibility of each method toward the Illumina spike-in concentration}\label{Tab1}
\resizebox{5cm}{!}{
\begin{tabular}{|l|r|r|}
\hline
Models	&	RMSE	&		K-L	\\ \hline
ENr	&	1.346	&		51,310		\\ \hline
ENn	&	1.407	&		41,010		\\ \hline
ENm	&	1.483	&		23,170		\\ \hline
ENmc	&	1.483	&		23,170		\\ \hline
EGm	&	1.470	&		20,660		\\ \hline
GN	&	1.521	&		58,480		\\ \hline
ELNn	&  1.411	&		41,200		 \\ \hline
ELNm	&	1.489&		21,280		\\ \hline
ELNp	&	1.423	&		37,800		\\ \hline
GLNn	& \textbf{1.323}	&		\textbf{4,333}		 \\ \hline
GLNm	&	1.510	&		29,630		\\ \hline
GLNp	&	10.700	&		-115,400		\\ \hline
\end{tabular}
}
\end{center}
\end{table}

\begin{table}[!h]
\begin{center}
\caption{Reproducibility of each method toward the Illumina spike-in based on the experiment data}\label{Tab2}
\resizebox{5cm}{!}{
\begin{tabular}{|l|r|r|}
\hline
Models	&	RMSE	&		K-L	\\ \hline
ENr	&	7.251	&		1,141,000	\\ \hline
ENn	&	7.127	&		106,2000	\\ \hline
ENm	&	6.927	&		926,500	\\ \hline
ENmc	&	6.927	&		926,200	\\ \hline
EG	&	6.919	&		907,900	\\ \hline
GN	&	7.100	&		1,183,000	\\ \hline
ELNn	&  7.124	&		1,062,000	\\ \hline
ELNm	&	6.904		&	911,600	\\ \hline
ELNp	&	7.092	&		1,035,000	\\ \hline
GLNn	& \textbf{6.825}	&	\textbf{793,400}	 \\ \hline
GLNm	&	6.937		&	968,400	\\ \hline
\end{tabular}
}
\end{center}
\end{table}

Table \ref{Tab1} shows that the GLNn reproduces the Illumina concentration better than others. The ENr shows the closest performance toward the GLNn. Here we can see that the GN method does not perform optimally. We also notice that the GLNp provides negative values for the Kullback-Leibler coefficient, therefore, this method is excluded from further comparisons. The behavior of GLNp which is different from other models, also shown at the supplemental plots.

Table \ref{Tab2} shows how each method reproduces the data from the experiment. We see that GLNn can be considered to reproduce it better than others, based on the RMSE, and the Kullback-Leibler coefficient.

Tables \ref{Tab1} and \ref{Tab2} provide insight about how the performance comparison among the models would be conducted further.

First, we compute the adopted Affycomp benchmarking criteria, based on the data after background correction and their log transformation.

Second, in the simulation, the $MSE_{bc}$ and the $L_{1}$ error will be computed based on the log transformation of the experiment and the nominal concentration data.

The log transformation that we use in this paper, respectively, for the benchmarking and the FFPE data sets are as follows
\begin{equation}
y=\log(x+\sqrt{(x^{2}+1)},\text{base}=2) \quad \text{and} \quad y=\log(x+1+\sqrt{(x^{2}+1)},\text{base}=2)
\end{equation}
 where $x$ is the concentration or the intensity value.

\subsubsection{Non-simulation} \label{sec711}

In Table \ref{Tab3}  it is shown that the ENr provides the smallest variation and IQR and the GLNn model provides the smallest 99.9\% percentiles of log fold change for the non spike-in between replicates. The largest variation, IQR, and 99.9\% percentiles, respectively are the GLNm, the ELNm and the GN.

\begin{table}[!h]
\begin{center}
\caption{Median SD, IQR and 99.9\% percentiles of log fold change for non spike-in between replicates for each model.}\label{Tab3}
\resizebox{9cm}{!}{
\begin{tabular}{|l|c|c|c|}
\hline
Aspects/Methods	&	Median SD &IQR	&	99.90\%	\\ \hline
ENr	&	\textbf{0.027}&\textbf{0.062}	&	0.415	\\ \hline
ENn	&	0.043& 0.089	&	0.441	\\ \hline
ENm	&	0.069& 0.139	&	0.486	\\ \hline
ENmc	&	0.069&0.139	&	0.486	\\ \hline
EGm	&	0.065& 0.134	&	0.477	\\ \hline
GN	&	0.051& 0.098	&	0.520	\\ \hline
ELNn	&	0.045&0.093	&	0.442	\\ \hline
ELNm	&	0.071& 0.145	&	0.489	\\ \hline
ELNp	&	0.049& 0.100	&	0.449	\\ \hline
GLNn	&	0.038& 0.075	&	\textbf{0.398}	\\ \hline
GLNm	&	0.076& 0.080	&	0.507	\\ \hline
\end{tabular}
}
\end{center}
\end{table}

\begin{table}[!h]
\begin{center}
\caption{The signal detect $R^{2}$ by regressing the Nominal and observed value for each model for the Illumina spike-in.}\label{Tab5}
\resizebox{9cm}{!}{
\begin{tabular}{|l|c|r|r|r|} \hline
Models	& Signal detect $R^2$ &	Low.$R^{2}$	& Med.$R^{2}$&	 High.$R^{2}$ \\ \hline
ENr& 0.959&0.618&\textbf{0.698}&\textbf{0.559}\\ \hline
ENn&0.958&0.622&0.695&0.557	\\ \hline
ENm&0.957&0.635&0.695&0.558	\\ \hline
ENmc&0.957&0.635&0.695&0.558	\\ \hline
EGm	& 0.957&0.633&0.695&0.558	\\ \hline
GN	&0.956&\textbf{0.650}&0.697&0.555	\\ \hline
ELNn&0.958&0.624&0.695&0.557	\\ \hline
ELNm&0.957&0.636&0.694&0.558 \\ \hline
ELNp&0.958&0.627&0.695&0.557	\\ \hline
GLNn&\textbf{0.960}&0.609&0.696&0.558\\ \hline
GLNm&0.956&0.637&0.694&0.558\\ \hline
\end{tabular}
}
\end{center}
\end{table}

In Table \ref{Tab5}, it is shown that, in general, all methods perform similar to each others.
The GLNn models have the highest signal detect $R^{2}$. The GN model has the highest $R^{2}$ at low concentration but has the lowest $R^{2}$ at high concentration. This means that the GN model works better at low concentration. On the other hand the ENr shows that it works better at medium and high concentrations, which is followed closely by GLNn model.

If we divide the concentrations into two categories, where  high concentration means that the nominal concentration is at least 3pM and low concentration means that the nominal concentration is at most 1pM, the GLNn model has the highest $R^{2}$ (the data is not shown here). It means, in general and at high concentrations, the GLNn offers the best fitted than other models.

As in Table \ref{Tab5}, Table \ref{Tab6} shows that all models have similar performance, although the GLNn model has the highest $R^{2}$ of nominal concentration against observed log-fold-change.

\begin{table}[!h]
\begin{center}
\caption{The $R^{2}$ observed log-fold-change against nominal log-fold-changes for the spike in genes.}\label{Tab6}
\resizebox{8cm}{!}{
\begin{tabular}{|l|c|c|}\hline
Methods	&	Obs-intended-fc.$R^{2}$	&	Obs-(low) int-fc.$R^{2}$	\\ \hline
ENr	&	0.976	&	0.989	\\ \hline
ENn	&	0.974	&	0.990	\\ \hline
ENm	&	0.972	&	0.985	\\ \hline
ENmc	&	0.972	&	0.985	\\ \hline
EGm	&	0.972	&	0.986	\\ \hline
GN	&	0.970	&	0.987	\\ \hline
ELNn	&	0.974	&	0.990	\\ \hline
ELNm	&	0.972	&	0.985	\\ \hline
ELNp	&	0.973	&	0.990	\\ \hline
GLNn	&	\textbf{0.978}	&	\textbf{0.991}	\\ \hline
GLNm	&	0.971 &	0.984\\ \hline
\end{tabular}
}
\end{center}
\end{table}

\begin{table}[!h]
\begin{center}
\caption{The AUC value for each model.}\label{Tab7}
\resizebox{13cm}{!}{
\begin{tabular}{|l|c|c|c|c|c|}
\hline
Methods	&	Low AUC &Med AUC	&	High AUC & Average AUC	 &All\\ \hline
ENr	&	0.450&0.987	&	\textbf{0.785} &0.585& 0.886	\\ \hline
ENn	&	0.518& 0.987	&	0.764 & 0.631& 0.899	\\ \hline
ENm	&	0.573& 0.987	&	0.741 & 0.667	& 0.911\\ \hline
ENmc	&	0.573& 0.987	&	0.741 & 0.667& 0.911	\\ \hline
EGm	&	0.567& 0.987	& 0.746 &	0.664& 0.910	\\ \hline
GN	&	0.552& 0.987	&	0.723 & 0.651& 0.904	\\ \hline
ELNn	&	0.524&0.987	&	0.763 &0.635& 0.900	\\ \hline
ELNm	&	0.574& 0.987	&	0.741 & 0.668& 0.912	\\ \hline
ELNp	&	0.534& 0.987	&	0.761 & 0.642& 0.902	\\ \hline
GLNn	&	0.498& \textbf{0.987}	&	0.784 &0.619&0.896	\\ \hline
GLNm	&	\textbf{0.579}& 0.987	&	0.730 & \textbf{0.671} & \textbf{0.913}	\\ \hline
\end{tabular}
}
\end{center}
\end{table}

Table \ref{Tab7} provides the results from the computation of the AUC value. The table shows that all models have a better accuracy at medium concentrations than at low and high concentrations.  The ENr, is followed by the GLNn, performs very poor at the low concentrations, but the GLNm is perform at best. At high concentrations, the ENr performs the best and it is followed by the GLNn. But in general, the highest AUC is achieved by all the model with the MLE parameter estimation methods: the GLNm, ELNm, and ENm.

The computation, which is based on the 12 and all arrays, provides the results where all models have the AUC greater than 0.9. According to Zhu et al. \cite{Zhu10}, the AUC between 0.9 and 1.0 is classified as excellent in measuring the accuracy. Therefore, based on Table \ref{Tab7}, we can identify that there are some models excellency accurate in predicting the gene expression.

In the Appendix, we put some supplementary materials, based on the Affycomp criteria from \cite{Cop04} and \cite{Iri03a,Iri03b} as explained in Section \ref{sec5} .

The MA plots \ref{figa1}, \ref{figa2} and \ref{figa3} show that all models perform similarly, except the GLNp model. In variance across replicates (Figures \ref{figb1} and \ref{figb2}), the GLNn model performs better than other models at the low and medium concentrations. At high concentrations the EGm and the GN models perform not at best. The computation (not showed) also provides the result that the GN model produces more differentially expressed non-spike in genes.

 A slight over-estimation is shown in figures \ref{figc1} and \ref{figc2}, where all models are above the ideal line at low and medium concentrations, particularly the GN and GLNm models, and then gradually goes under the ideal line at high concentrations.

\subsubsection{Simulation} \label{sec712}

We do simulations to assess the performance of each model. The bias of the background correction is assessed by the MSE$_{bc}$, and the bias of the parameterizaton is assessed by $L_{1}$ error. These criteria have been defined at Section \ref{sec6}.

\begin{table}[!h]
\begin{center}
\caption{The simulation results on Spike in data set.}\label{Tab8}
\resizebox{11cm}{!}{
\begin{tabular}{|l|c|c|c|c|c|c|}
\hline
Aspects	&	MSE$_{bc}$	&	\multicolumn{4}{|c|}{$L_{1}$ error	 }						\\ \hline
Methods	&		&	$\alpha$	&	$\beta$	&	$\mu$	&	 $\sigma$	\\ \hline
ENr	&	0.0451	&		&	0.6638	&	46.58	&	11.44	\\ \hline
ENn	&	0.0490	&		&	0.6252	&	41.61	&	2.806	\\ \hline
ENm	&	0.0383	&		&	0.6101	&	58.92	&	2.04	\\ \hline
ENmc	&	0.0362	&		&0.6104		&	62.77	&	2.039	 \\ \hline
GN	&	\textbf{0.0297}	&	\textbf{0.000336}	&	 \textbf{0.007332}	&	0.012920	&	\textbf{0.0149}	\\ \hline
ELNn	&	0.0484	&		&	0.009082	&	0.000427	&	 0.0177	\\ \hline
ELNm	&	0.0393	&		&	0.840300	&	0.000402	&	 0.0180	\\ \hline
ELNp	&	0.0605	&		&	0.472000	&	0.000429	&	 0.0180	\\ \hline
GLNn	&	0.2156	&	0.051750	&	0.055060	&	 0.000427	&	0.0176	\\ \hline
GLNm	&	84.3700	&		38.860000	&	0.851100	&	 \textbf{0.000354}	&	0.0169	\\ \hline
\end{tabular}
}
\end{center}
\end{table}

From the simulation results in Table \ref{Tab8} we can see that simulation results of the EG model are not available, because the MBCB \textit{package} could not work at  the log transformation that we have chosen. The GN model performs best, by providing the smallest bias for the background correction and the parameters. A close performance is achieved by the ELN, particularly ELNn. The GLNn does not have an optimal performance on the MSE$_{bc}$, but we still can consider its performance good, concerning that the bias of the parameters are similar to other proposed models and GN.

One of the proposed models, GLNm has the highest bias on the MSE$_{bc}$ and the parameter $\alpha$. In our view this  happens because we use an approximation in estimating the true intensity value. The EN models (ENr, ENm, ENn and ENmc) have considerably better performance at $MSE_{bc}$, but are not good at the parametrization. The bias on the parametrization of the noise is higher than in other models.

\subsection{FFPE data set} \label{sec72}

Based on the results from Section \ref{sec71}, we compare the performance of these models on the public data sets. We would like to know how good these models are
in real data samples. Here, we choose to use the FFPE data set.

Currently the FFPE archival samples are widely available in million and it is a great source of  information in medical studies about some diseases, for example cancer. This data type is suffering from the RNA degradation, which leads to poor performance in array-based studies. However, the Illumina's DASL assays could provide high-quality data from this degraded RNA samples.

Comparing the performance of these background correction models certainly would help researchers to choose the appropriate background correction for their data, particularly if their data is the FFPE type.

The background correction for the FFPE data set is implemented in three steps:
\begin{description}
\item[\emph{step 1}] Do the quality control (QC) to the raw FFPE data. In this paper, we used the \emph{ffpe} package in R (\cite{Wal14}).
\item[\emph{step 2}] Do the data transformation ($\log ((x+1+\sqrt(x^{2}+1)),base=2)$) to the raw FFPE data after QC and estimate the BC parameters based on it. The estimators of true intensity value and the background correction are based on the regular and negative control probe intensity data respectively.
\item[\emph{step 3}] Compute the true intensity value (the adjusted intensity estimator) based on the BC parameters at \emph{step 2}.
\end{description}

The results of our computation are in Tables \ref{Tab9} and \ref{Tab10}. From these tables, we can see that there are no EG and GN models. Neither of these models can work on these data sets. For some samples in the data set, both models fail to compute the parameters which has the consequence that the true intensity value cannot be provided.

We decided to remove the EG and GN models from further comparisons in both FFPE data sets. Here we provide the results of the rest of the models only.

\begin{table}[!h]
\begin{center}
\caption{Simulation results on the GSE32651 data set.} \label{Tab9}
\resizebox{10cm}{!}{
\begin{tabular}{|l|r|r|r|r|r|r|}
\hline
Aspects	&	MSE$_{bc}$	&	\multicolumn{4}{|c|}{$L_{1}$ error	 }						\\ \hline
Methods	&		&	$\alpha$	&	$\beta$	&	$\mu$	&	 $\sigma$	\\ \hline
ENr	&	0.0581	&		&	0.6571	&	297.6705&	22.6090	\\ \hline
ENn	&	0.2810	&		&	0.5719	&	1.9840 &	2.6265	\\ \hline
ENm	&	0.0859	&		&	0.5911	&	28.0490	&	1.7150	\\ \hline
ENmc	&	\textbf{0.0362}	&		&	0.6104	&	62.7700	&	 2.0390	\\ \hline
ELNn	&	0.2749	&		&	\textbf{0.0245}	&	0.0014	&	 0.0184	\\ \hline
ELNm	&	0.0587	&		&	0.8256	&	\textbf{0.0004}	&	 0.0184	\\ \hline
ELNp	&	0.6724	&		&	0.4873	&	0.0013	&	0.0179	 \\ \hline
GLNn	&	0.8378	&	\textbf{0.3401}	&	0.5273	&	0.0014	 &	0.0179	\\ \hline
GLNm	&	84.1500	&	71.8700	&	0.8867	&	0.0005	&	 \textbf{0.0175}	\\ \hline
\end{tabular}
}
\end{center}
\end{table}

\begin{table}[!h]
\begin{center}
\caption{Simulation results on the GSE32489 data set.} \label{Tab10}
\resizebox{10cm}{!}{
\begin{tabular}{|l|r|r|r|r|r|r|}
\hline
Aspects	&	MSE$_{bc}$	&	\multicolumn{4}{|c|}{$L_{1}$ error	 }						\\ \hline
Methods	&		&	$\alpha$	&	$\beta$	&	$\mu$	&	 $\sigma$	\\ \hline
ENr	&	\textbf{0.0929}	&		&0.6648		&	67.1700	&	 9.5110	\\ \hline
ENn	&	0.8629	&		&	0.5086	&	0.9359	&	2.0391	\\ \hline
ENm	&	0.1817	&		&	0.5580	&	14.1967	&	1.7117	\\ \hline
ENmc	&	0.1840	&		& 0.5564 		&	14.1789	&	 1.0490	\\ \hline
ELNn	&	1.0545		&	&0.8574		&	0.0018	&	0.0182	 \\ \hline
ELNm	&	0.1161	&		&0.7809		&	0.0006	&	0.0177	 \\ \hline
ELNp	&	1.2471		&		&0.4614		&	0.0018&	0.0177	 \\ \hline
GLNn	&	1.3476		&	\textbf{0.3320}	&	\textbf{0.4974}	 &	0.0018	&	\textbf{0.0175}	\\ \hline
GLNm	&	164.9800		&	22.2390	&	0.8048	&	 \textbf{0.0004}	&	\textbf{0.0175}	\\ \hline
\end{tabular}
}
\end{center}
\end{table}

Tables \ref{Tab9} and \ref{Tab10} consistently show that the bias of the parameters of noise in the EN models are higher than the proposed models. For the parameter $\beta$, the ELNn has the smallest bias and it is followed by the ELNp and the GLNn. With regard to the bias of the background correction, the EN models show the smallest bias in both of FFPE data sets.

The proposed model GLNm continues to show the highest bias in the background correction and the parameter $\alpha$. As it has been mentioned previously, this is a consequence of the approximative computation of the true intensity value, where we take the approximation until $k=10$. There is a possibility to apply different numerical approximation for both of the ELN and GLN models.

\section{Conclusions and indication of future work}\label{sec8}

We have studied the additive models of background correction for beadarrays and proposed some new models where the true intensity is assumed to have exponential or gamma distribution and the noise is lognormally distributed. We have derived the estimator of the true intensity value of the proposed models.

Further, we compared the performance of all models, based on the benchmarking and public data sets. In the benchmarking data set we adopted the criteria from the Affycomp \cite{Cop04} and for the simulation study we used the criteria which have been used in \cite{Xie09}, \cite{Che11} and \cite{Pla11,Pla12}. For the public data sets, we only used the criteria for the simulation study.

We have seen in Sections \ref{sec711} and \ref{sec712} that EN, EG, GN and GLN perform rather similarly. From the affycomp criteria we can provide the following points: \begin{enumerate}
\item{the ENr and GLNn provide the lowest variation between replicates and all models using the MLE estimation method have a higher variation than others}
\item{the GLNn model has the highest signal detect $R^{2}$ in general and in high concentration. This means the GLNn model is the best fitted for the gene expression.}
\item{the GLNn model, based on the MvA plot, produces the least number of genes which should not be expressed but  are nevertheless expressed. On the other hand, the GN model provides the largest number of such genes.}
\item{all models with the MLE estimation method have a higher average AUC value, which means that they provide a better accuracy in predicting the gene expression.}
\end{enumerate}

In the simulation study, the best performance in the background correction and parametrization error is achieved by the GN model. It is followed by our proposed ELN models. It has been shown that the GLNn does not perform optimally at the MSE$_{bc}$ criterion, but for the parametrization this model still can be considered good.

In the FFPE public data set, the GN and EG models cannot be implemented. This is in strong contrast with the fact that in the simulation study of benchmarking data set, the GN model has the best performance.

The EN models show the highest bias in the parametrization in both public data sets and the lowest bias in the background correction. Our proposed models, except the GLNm, show the lowest bias in the parametrization in both data sets and a moderate bias in the background correction.

Based on the results from the benchmarking data and the public data sets, we would suggest researchers the following:
\begin{enumerate}
\item if the GN model works properly at the data set at hand, then use the GN model to correct the background.
\item if the GN model fails, then use our proposed models, particularly the GLNn model. The reason for not choosing the ELN models is that the value of the parameter $\alpha$ from the benchmarking data set is less than 1, around 0.2. Therefore, the gamma model is more appropriate to model the true intensity distribution than the exponential one. We believe that the right approximative computation of the GLN models will lead to a better performance than the current approximation.

The ELN models perform better than the original EN models, due to the fact that not only the regular probes, but also the control probes are skew-distributed  (\cite{Che11}). Therefore, these models could be the second choice after the GLN, when the GN model does not work.
\item With regard to the computation time, at the benchmarking data set the EN models are working faster than the others. They are followed by the ELNp, ELNn, and EGm. The GLNn and the ENmc are the third fastest, then come the GN and the ELNm, which are followed by the GLNm as the slowest one.

\item As we will show in a subsequent paper, it is  possible to develop a new model which satisfies all of the affycomp criteria, the consistency in the background correction and the parametrization errors.
\end{enumerate}

One of the purposes of using microarray technology is finding the genes which are expressed differentially due to some disease or condition. Therefore, it is important to investigate the effect of bias of the background correction and the parametrization toward the differentially expressed genes. This would be our future work.

\textbf{Acknowledgement.}
The author would like to thank \textit{Prof.\ Istv\'{a}n Berkes} for guidance in writing this work, the Austrian Science Fund (FWF), Project P24302-N18 for financial support and \textit{Levi Waldron, Ph.D} and \textit{Prof.\ Ernst Stadlober} for their comments. We also thank an anonymous referee for valuable remarks leading to an improvement of the paper.

\bibliographystyle{plainnat}
\bibliography{emaRef}

\newpage{}
\begin{appendices}

\section{MA plots}
\begin{figure}[!h]
     \centering
     \includegraphics[width=6in, height=7in]{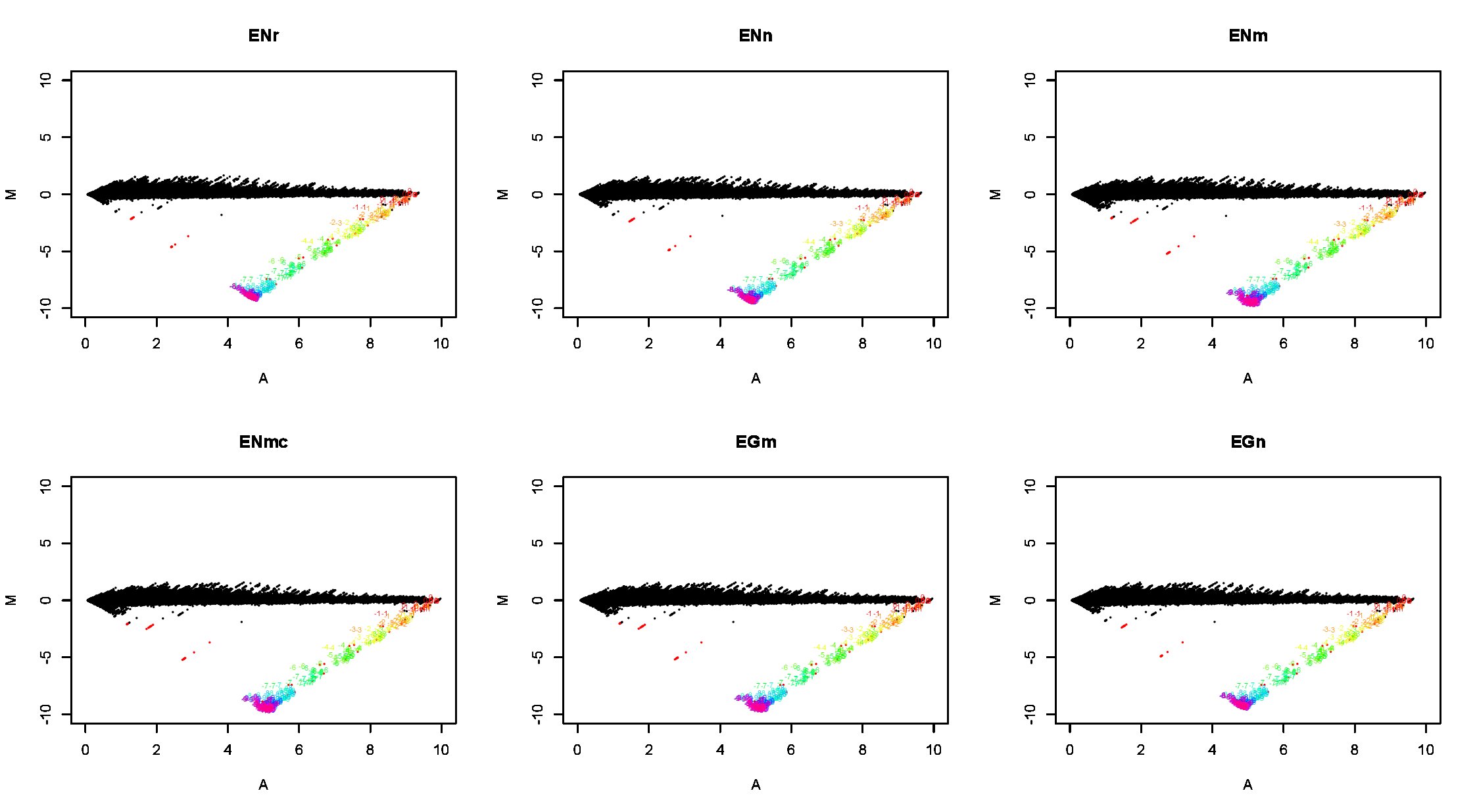}
     \caption{MA plots. \emph{(cont.)}} \label{figa1}
  \end{figure}

\begin{figure}[tbp]
     \centering
    \includegraphics[width=6in, height=8in]{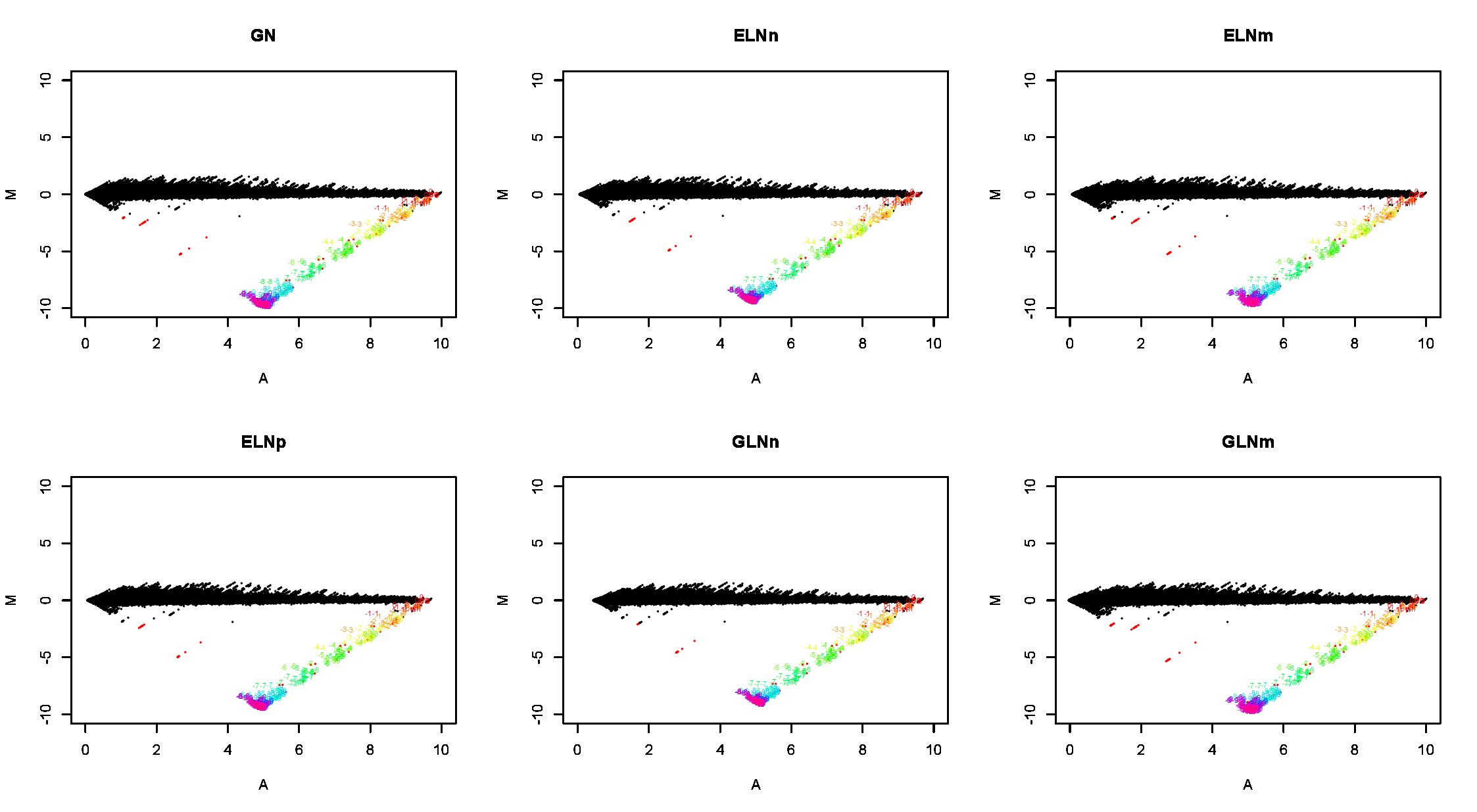}
     \caption{MA plots. \emph{(cont.)}} \label{figa2}
  \end{figure}

\begin{figure}[t]
     \centering
     \includegraphics[width=2.75in, height=3in]{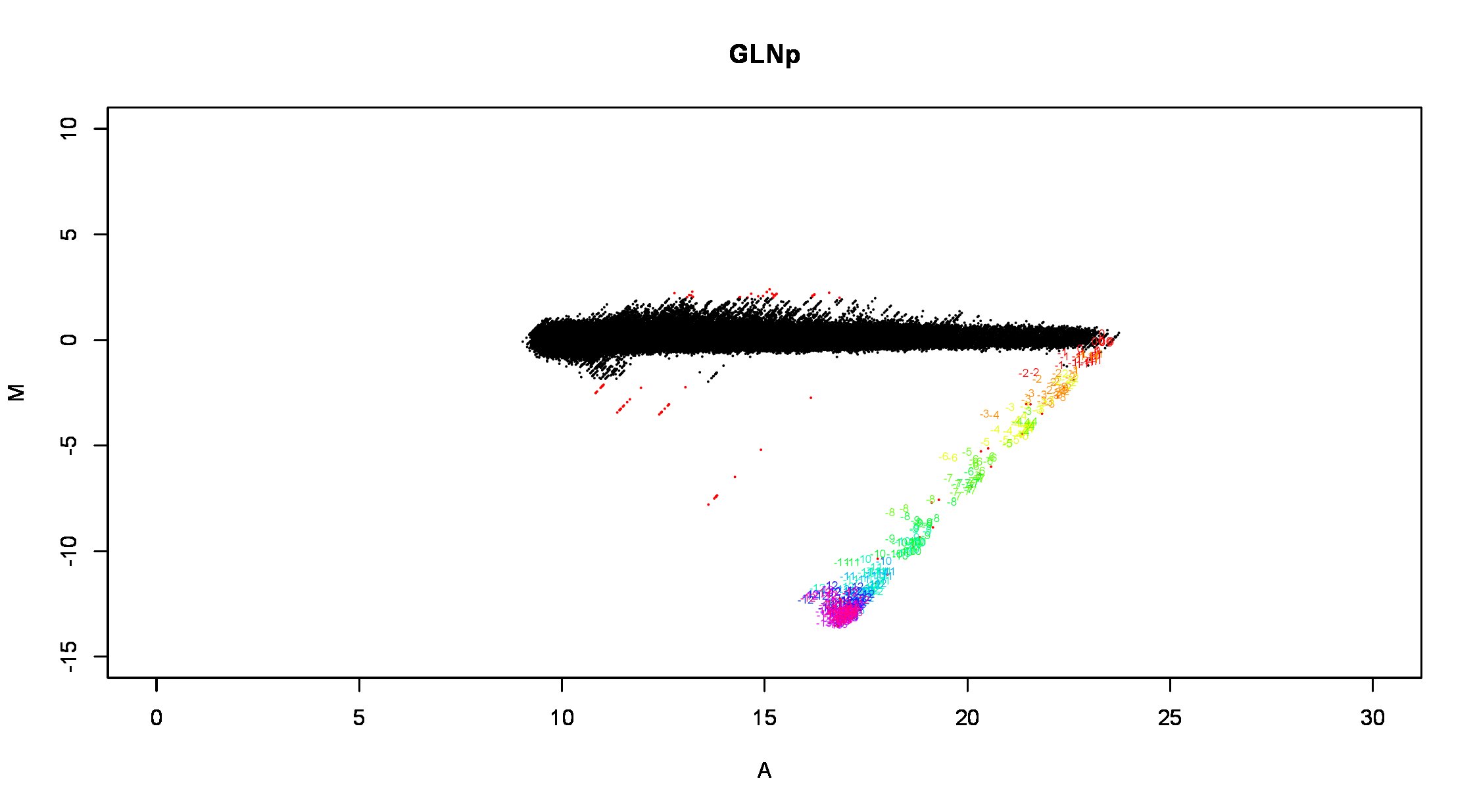}
     \caption{MA plots.} \label{figa3}
  \end{figure}

\clearpage
\section{Variance across replicates}
\begin{figure}[!h]
     \centering
     \includegraphics[width=3.4in, height=3.4in]{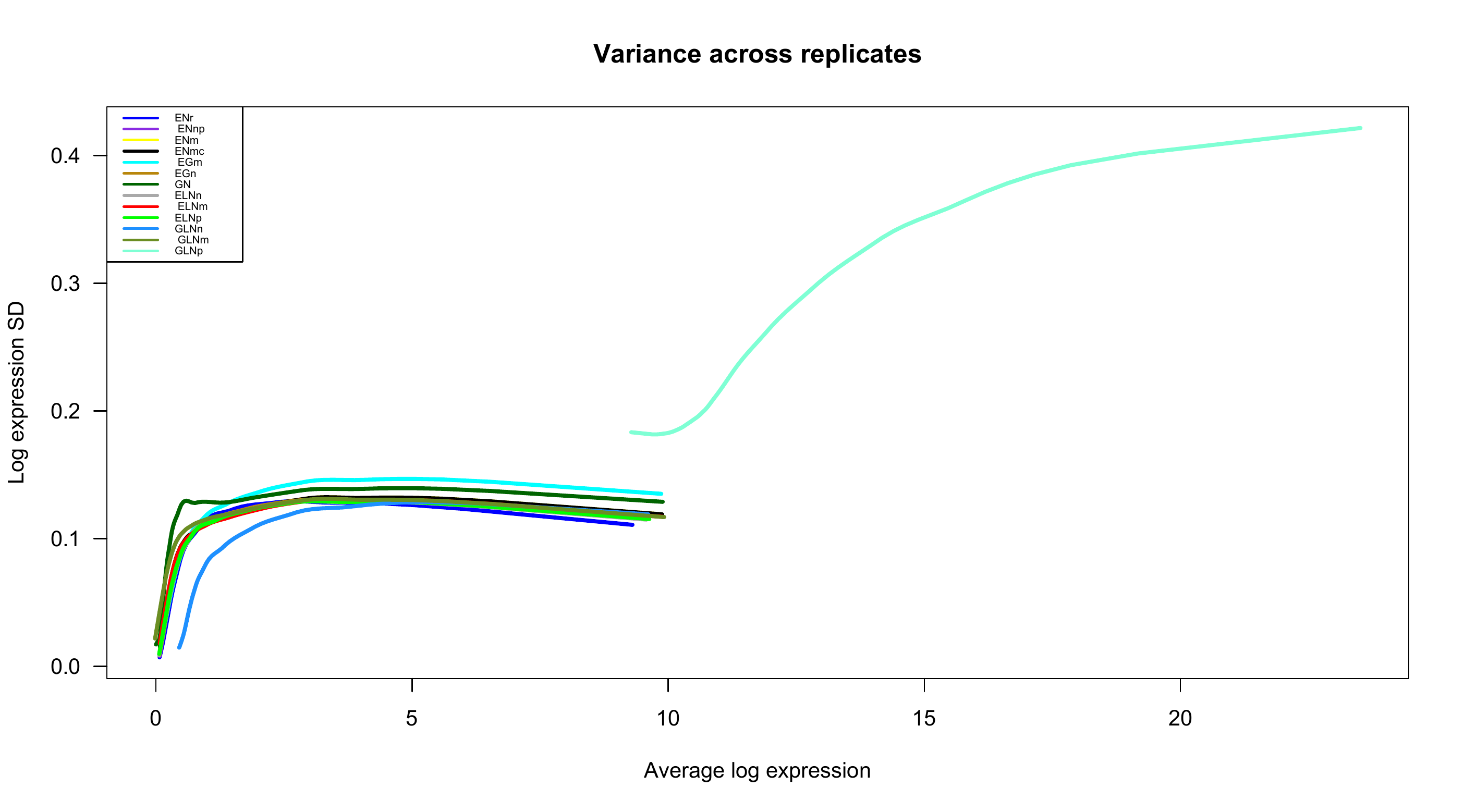}
     \caption{Variance across replicates plots, all models} \label{figb1}
  \end{figure}

\begin{figure}[!h]
     \centering
     \includegraphics[width=3.4in, height=3.4in]{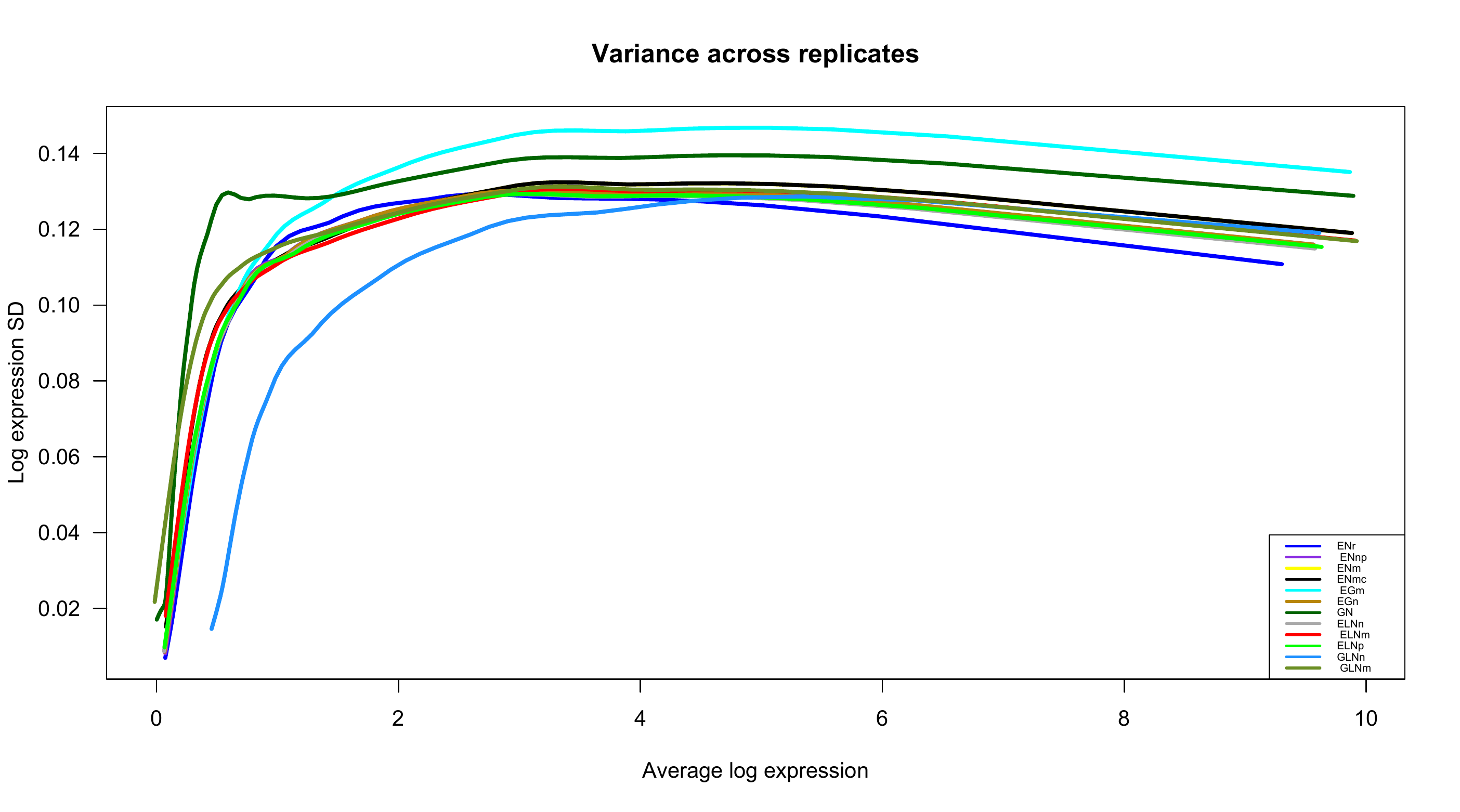}
     \caption{Variance across replicates plots, without GLNp. } \label{figb2}
  \end{figure}

\clearpage
\section{Nominal vs Observed intensity}
\begin{figure}[!h]
     \centering
     \includegraphics[width=3.4in, height=3.4in]{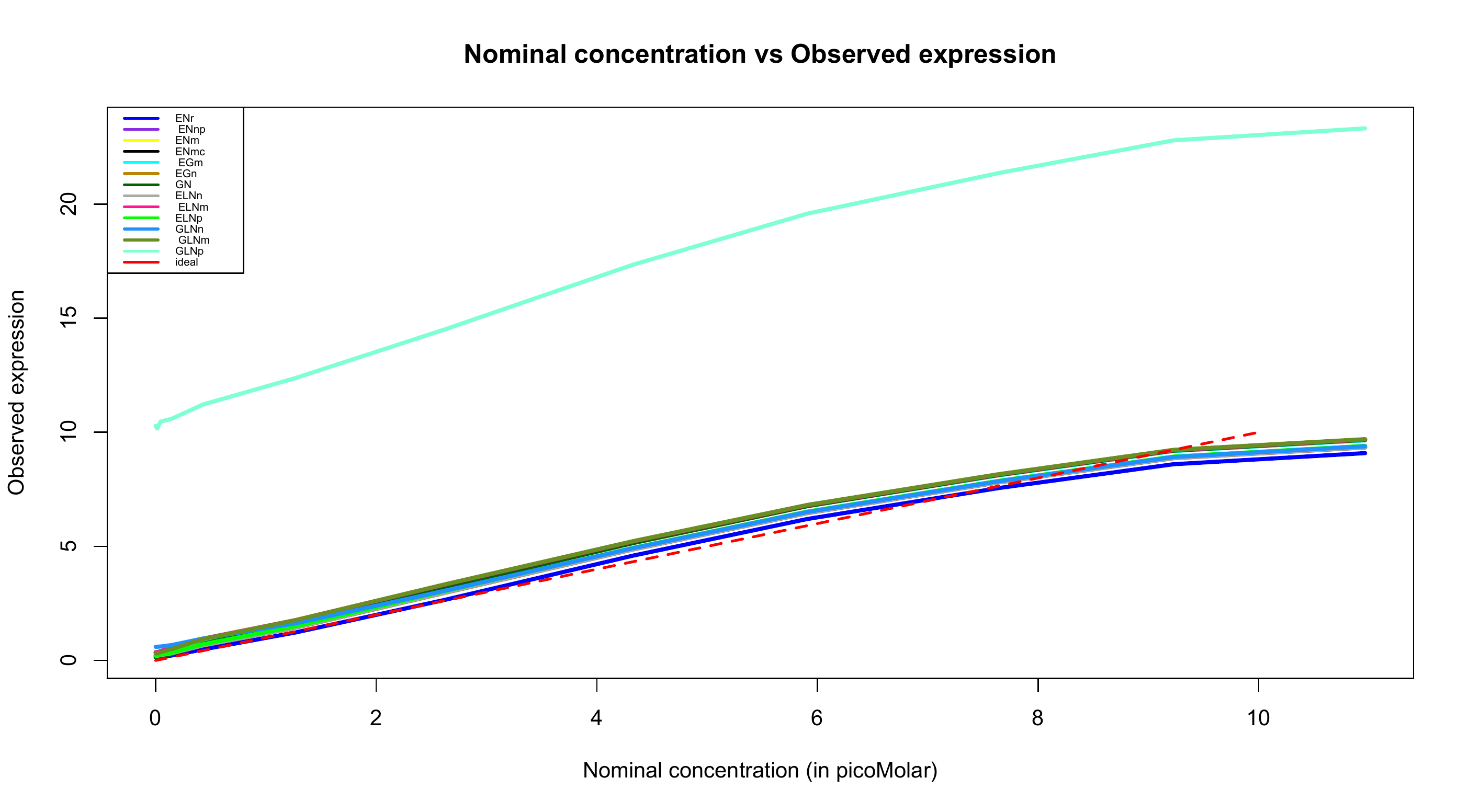}
     \caption{Observed vs Nominal plots, all models} \label{figc1}
  \end{figure}

\begin{figure}[!h]
     \centering
     \includegraphics[width=3.4in, height=3.4in]{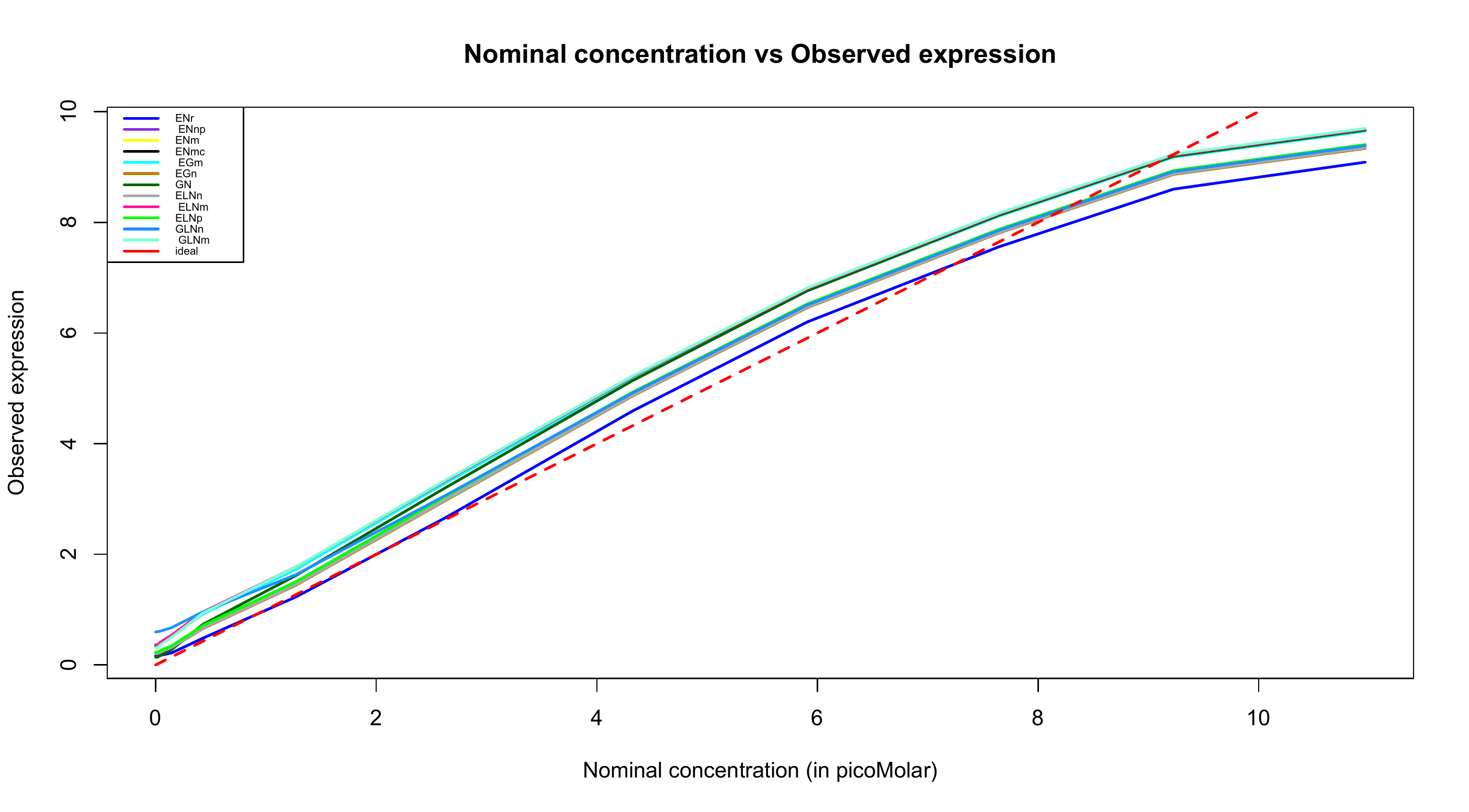}
     \caption{Observed vs Nominal,  without GLNp.} \label{figc2}
  \end{figure}

\clearpage
\section{Nominal vs Observed fold change plots}
\begin{figure}[!h]
     \centering
     \includegraphics[width=3.4in, height=3.4in]{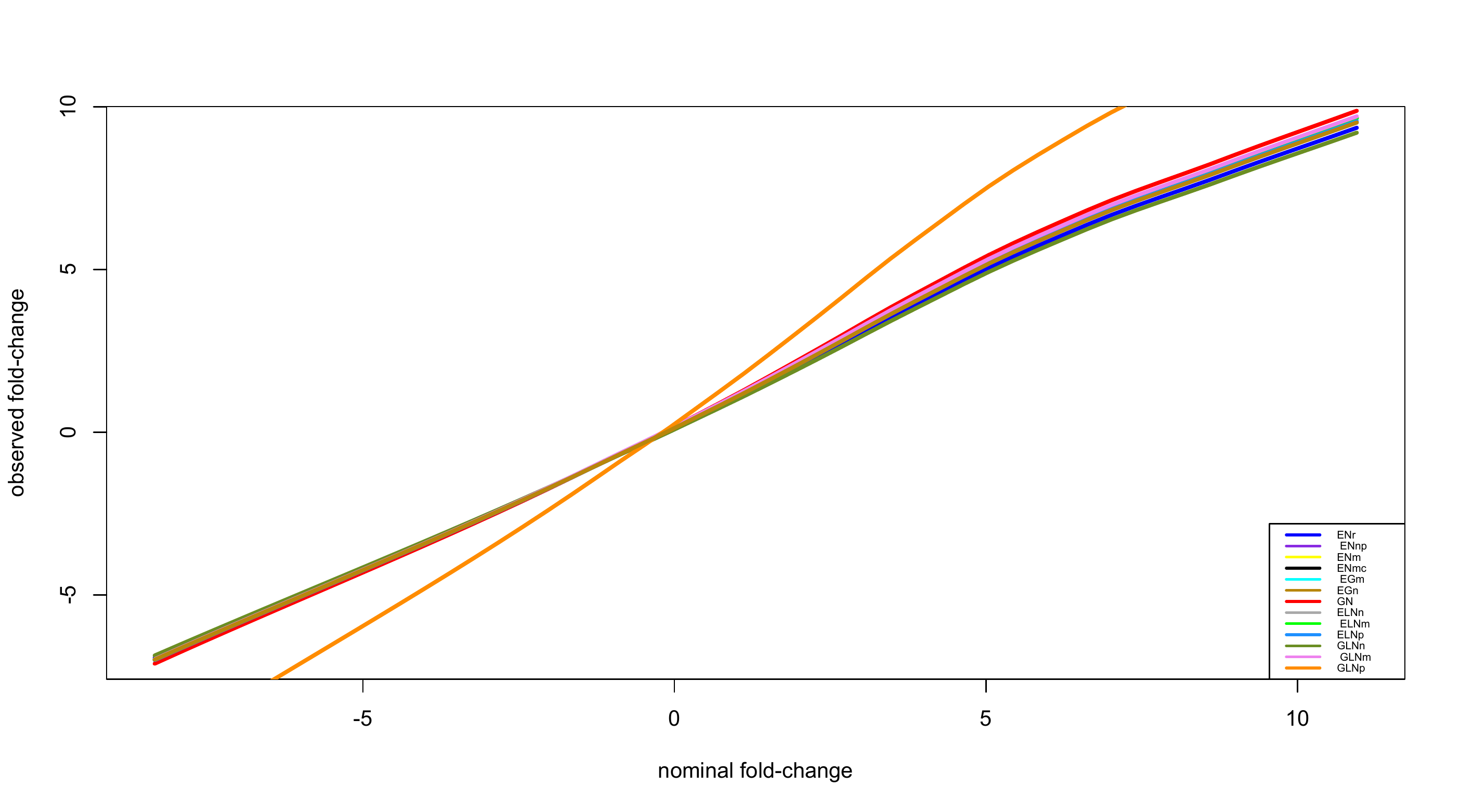}
     \caption{Observed vs Nominal log fold-change plots, all arrays} \label{figd1}
  \end{figure}

\begin{figure}[!h]
     \centering
     \includegraphics[width=3.4in, height=3.4in]{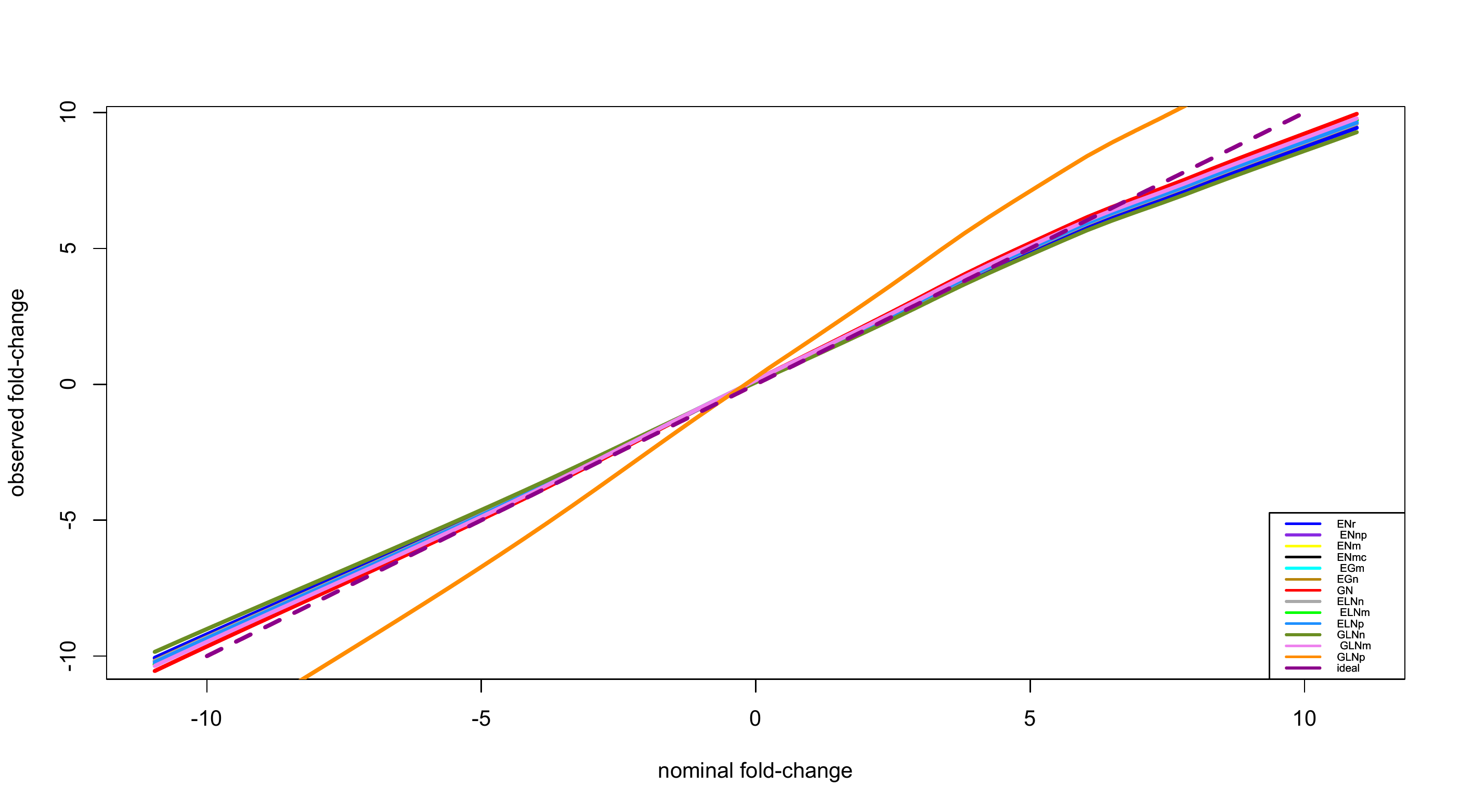}
     \caption{Observed vs Nominal log fold-change plots, 24 arrays.} \label{figd2}
  \end{figure}

\clearpage
\section{ROC curves}
\begin{figure}[!h]
     \centering
     \includegraphics[width=3.4in, height=3.4in]{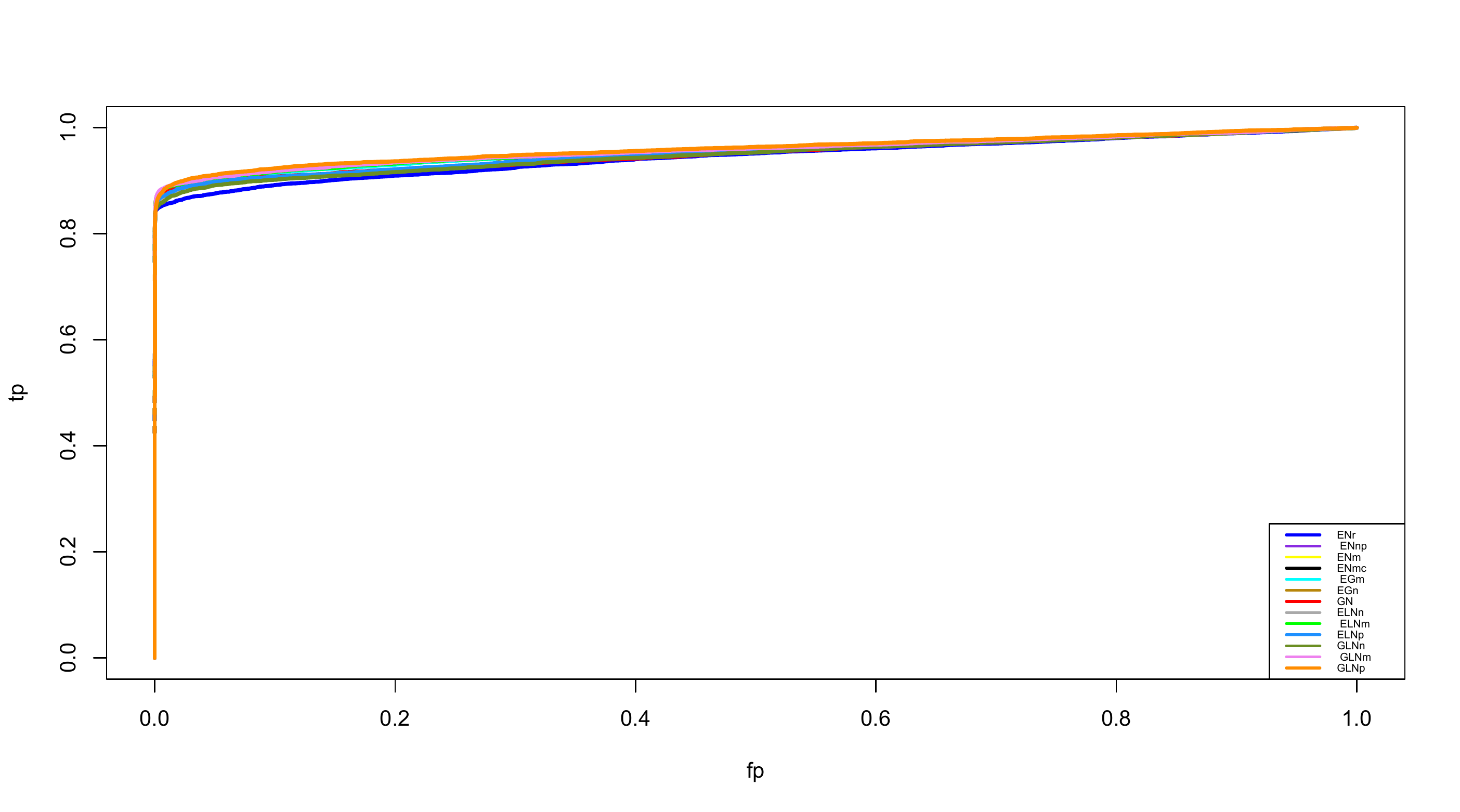} \label{fige1}
     \caption{ROC plots, all models}
  \end{figure}
\begin{figure}[!h]
     \centering
     \includegraphics[width=3.4in, height=3.4in]{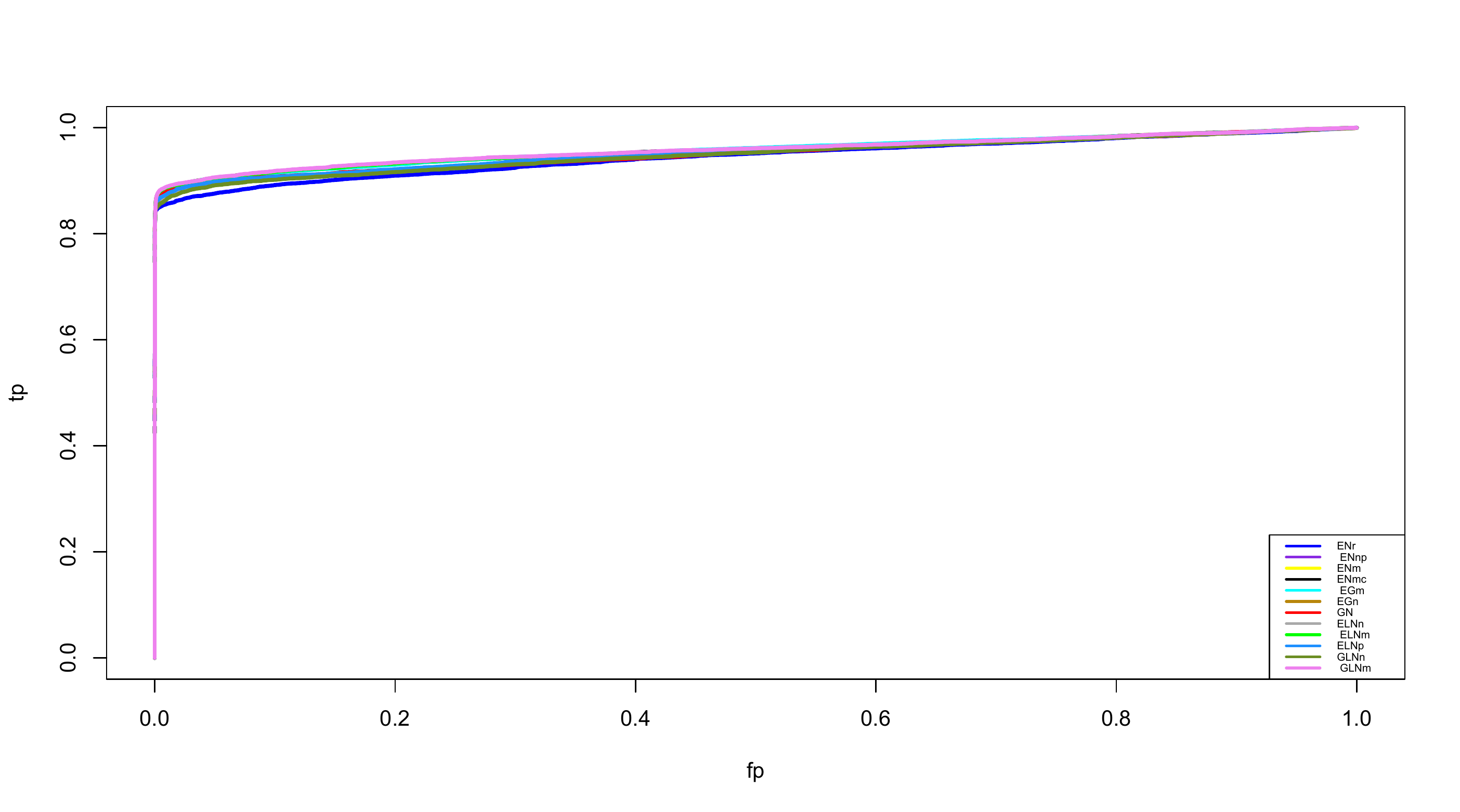} \label{fige2}
     \caption{ROC plots without GLNp}
  \end{figure}

\end{appendices}

\end{document}